\documentclass[reprint, amsmath, amssymb, aps, longbibliography]{revtex4-2}

\usepackage{graphicx}
\usepackage{booktabs}
\usepackage{xcolor}
\usepackage{hyperref}

\begin{document}

\title{The Forgotten History of Wave Function Variance Optimization and its Relevance for Neural-Network VMC}

\author{Dario Bressanini}
\email{dario.bressanini@uninsubria.it}
\affiliation{Dipartimento di Scienza e Alta Tecnologia, Università dell'Insubria, Como, Italy}

\date{\today}

\begin{abstract}

We revisit the almost century-old question of which functional of the local energy best optimizes a trial wave function, a problem of central importance in Variational Monte Carlo (VMC) and, more recently, in Neural-Network VMC (NN-VMC). While variance optimization dates back to the 1930s, the high statistical noise and heavy-tailed local energy distributions inherent to modern neural-network wave functions have renewed interest in this approach. We retrace its long and largely forgotten history here, showing its direct relevance to modern Neural Quantum States (NQS) frameworks. Minimizing the variance (an $L^2$ norm) implicitly assumes a Gaussian local energy distribution: an unjustified assumption. For Coulombic systems, the local energy distribution exhibits $E^{-4}$ power-law tails, causing the Central Limit Theorem to fail for the variance estimator. This instability can be mitigated by robust cost functions: the Mean Absolute Deviation (MAD, an $L^1$ norm), the Cauchy loss, or the $\mathcal{L}_{-4}$ functional, which features a tail analytically designed to match the $E^{-4}$ exponent. We benchmark these functionals on $\mathrm{H}_2^+$, an exactly solvable system at every internuclear distance, using the Guillemin–Zener wave function across the full potential energy curve. While energy minimization by construction yields the lowest energy, variance minimization is surpassed at every $R$ by alternative functionals: MAD proves superior in the bonding region, while $\mathcal{L}_{-4}$ performs best in the dissociation regime.

\end{abstract}

\maketitle

\section{Introduction}

Trial wave function optimization in variational quantum Monte Carlo (VMC) is usually presented as a choice between minimizing the variational energy $E=\langle H\rangle$ and minimizing the variance of the local energy, 
\begin{equation}
\sigma^2 = \langle H^2\rangle - \langle H\rangle^2,  
\end{equation}
 where the local energy is defined as $E_L=H\Psi/\Psi$. The variance $\sigma^2$ is zero if and only if the trial function is an exact eigenstate, so minimizing it over a fixed sample of configurations gives a well-defined, numerically tractable target, even for excited states. After Umrigar, Wilson, and Wilkins put this ``zero-variance principle'' on a firm computational footing in 1988, variance minimization became the workhorse of VMC optimization for over a decade \cite{umrigar1988}.

During the 2000s, however, it slowly fell out of fashion, overshadowed by algorithms that minimize the energy directly: Stochastic Reconfiguration (SR) \cite{Sorella2001} and the Linear Method (LM), introduced by Nightingale and Melik-Alaverdian \cite{NightingaleMelikAlaverdian2001} and later extended to general nonlinear wave function parameters \cite{Umrigar2007,ToulouseUmrigar2007}.

With the arrival of neural-network trial wave functions introduced by Carleo and Troyer \cite{CarleoTroyer2017} and developed into all-electron forms by FermiNet \cite{pfau2020}, PauliNet \cite{Hermann2020}, and their successors, there has been a renewed interest in alternative cost functions. Highly flexible functions trained by stochastic gradient descent are prone to occasional, very large local energy deviations: the distribution of $E_L$ is heavy-tailed, with extremes arising near nodal surfaces and Coulomb singularities, where $E_L$ generically diverges. A handful of such outliers in a finite ensemble can dominate a variance-based cost function and destabilize training.
Some implementations address the problem by integrating estimators based on robust statistics into the energy optimization process. These modern approaches often encounter the same problems that have plagued variance optimization since its inception, 90 years ago.

Unfortunately, the history of variance optimization is largely forgotten in modern computational literature. This paper has two purposes: to recover this history and to test, on a benchmark with an essentially exact solution, the alternative robust cost functionals suggested in the past. We hope both will be of use to the Neural Quantum State (NQS) community and to quantum Monte Carlo (QMC) practitioners at large.

Section~\ref{sec:history} retraces this history from its origins as a diagnostic tool in the lower-bound theory of the late 1920s; through its recasting as an optimization criterion in the 1930s, the discrete-point schemes of mid-century, and the first Monte Carlo methods; to the golden age of quantum Monte Carlo and its subsequent decline, culminating in the recent resurgence in Neural-Network VMC (NN-VMC).

Section~\ref{sec:theory} presents the theoretical background: the relationship between energy error and variance, the cost functionals constructed from a maximum-likelihood perspective, and their interpretation as fitting a trial potential.

Section~\ref{sec:methods} describes the $\mathrm{H}_2^+$ benchmark and the two ansatzes used to evaluate the five cost functions; the performance of these functionals across the full potential energy curve is then reported in Section~\ref{sec:results}.

Section~\ref{sec:conclusions} summarizes what the history and the benchmark jointly teach about the choice of a cost function, and Appendix~\ref{app:centering} discusses the choice of the centering energy $E_R$ and the exact evaluation of the cost function gradients.

\section{A History of Variance Optimization}\label{sec:history}

\subsection{The Early Uses of Variance}
The story begins in 1928 when the energy variance $\sigma^2$ was first used not as a target for minimization, but as a diagnostic tool to estimate the accuracy of an energy already obtained by the Rayleigh-Ritz (RR) method. Temple showed that if $\langle H\rangle$ and an estimate of the first excited-state energy $E_1$ are both known, then \cite{temple1928}
\begin{equation}
E_0 \;\ge\; \langle H\rangle - \frac{\sigma^2}{E_1-\langle H\rangle}, \qquad \langle H\rangle < E_1 .
\label{eq:temple}
\end{equation}
Two years later, in a note added in proof, Eckart \cite{Eckart1930} derived a relationship between the root-mean-square error $q$ of a normalized trial wave function and its energy error, where $q^2=\int (\Psi_0 -\Psi)^2\,d\mathbf{r}$ and $\Psi_0$ is the exact ground state wave function. He showed that the square of the wave function error is bounded from above by the energy error divided by the first excitation gap: $q^2\le (E - E_0)/(E_1 - E_0)$, which provides a practical upper limit on the distance of the trial function from the exact ground state.

Weinstein, whose papers make no reference to Temple, independently attacked the same problem. His 1932 paper derived a bound essentially identical to Temple's \cite{Weinstein1932}. Later the same year he introduced the bound that bears his name, $E_0 \ge \langle H\rangle - \sigma$, which requires no excited-state energy but is valid only provided $\langle H\rangle \le (E_0+E_1)/2$, a restriction he
considered almost always satisfied \cite{Weinstein1932b}. In 1934 he derived its exact symmetric generalization
\begin{equation}
|\langle H\rangle - E_j| \;\le\; \sigma
\label{eq:weinstein}
\end{equation}
where $E_j$ is some eigenvalue, not necessarily that of the target state \cite{Weinstein1934}. Weinstein also proposed minimizing the variance as a variational principle in its own right, arguing that it provides a better criterion for the ``best'' approximated wave function than the Rayleigh-Ritz upper-bound method. Furthermore, he pointed out that the formal identity $\langle H^2 \rangle=\int(H \Psi)^2d\tau$ can fail when the eigenfunction expansion fails to converge. This is probably the earliest warning of the local energy divergence problems. 

In that same year, MacDonald showed that a whole family of functionals can be constructed, bounding the spectrum via the $m$-th expectation value $\int\psi(H-E_R)^{m}\psi\,d\tau$ for an arbitrary reference constant $E_R$, with $m=1$ recovering the ordinary Rayleigh-Ritz result and $m=2$ recovering Weinstein's bound \eqref{eq:weinstein} \cite{macdonald1934}.

Stevenson in 1938 added a free parameter $\alpha$ to Weinstein's bound and proposed maximizing the lower bound for a well-chosen $\alpha$~\cite{stevenson1938b}. In a follow-up paper, Stevenson and Crawford applied the method to helium, using a nine-term Hylleraas-type function, observing the two-sided bracket of the ground state energy tighten as the number of terms increased \cite{Stevenson1938}.

The interest in variance as a tool to generate lower bounds continued for more than two decades: Fr\"oman and Hall in 1961 tested what they called the ``minimum width principle'' (minimizing the variance) along with the Temple and Weinstein bounds, on the hydrogen atom and three helium trial functions of increasing quality. They studied the effect of the introduction of a uniform scale factor into the wave function, an old trick used in early quantum chemistry calculations to force a trial wave function to satisfy the virial theorem. They found that which bound is tighter depends on the trial function quality, with cruder functions favoring Weinstein and better ones favoring Temple \cite{fromanhall1961}.
They also plotted $E$ against $\sigma$ as the scale factor was varied, providing an early instance of the $(\sigma^2,E)$ plots that would later be fully explored by Conroy \cite{Conroy1967} and used in modern NN-VMC extrapolations.

In all of this work, variance was used as a diagnostic, to bracket the ground state energy with a lower bound, once the upper bound, usually obtained by the RR method, was already available.

\subsection{Variance as an Optimization Target}

The shift from using variance to bound a previously computed energy to employing it as an optimization criterion in its own right, already suggested by Weinstein, was computationally explored by Bartlett, Gibbons, and Dunn in 1935 \cite{Bartlett1935}.
They proposed using a general least-squares procedure to evaluate how well any function satisfies a differential equation. To our knowledge, this represents the first application of this approach to the Schr\"odinger equation.
Writing $(H-E)\Psi=\epsilon\Psi$, they treated $\sigma^2=\int\Psi^2\epsilon^2\,d\tau$ as a ``mean square energy deviation'' and $\epsilon$ as a pointwise indication of how well the wave equation is satisfied locally. They applied this method to the six-term Hylleraas helium function and noted a practical limitation that persists in modern VMC: $\epsilon$ diverges at the Coulomb singularities. In a footnote they even suggested weighting the squared deviation to emphasize different regions of space.

In a 1937 paper titled ``Criteria of Goodness for Approximate Wave Functions'', James and Coolidge \cite{James1937} demonstrated that the energy error $\langle H\rangle - E_0$, the root-mean-square wave function error $q$, and the root-mean-square local energy deviation $\sigma$ are related but distinct figures of merit, and that a wave function optimized for one is not optimized for the others.
They showed that for a sufficiently good wave function, the variance $\sigma^2$ is approximately proportional to the energy error, a relation that has recently been exploited in neural-network VMC calculations \cite{Fu2024, Chen2024}. Furthermore, they observed that variance optimization systematically reduces short-range errors near the singularities, at the expense of long-range accuracy.

This study presents, as far as we have been able to determine, the first actual variance minimization (over a single shielding parameter of a simple helium wave function) and the first use of the term ``local energy''.

The following year, James and Yost performed a direct quantitative test on the $1s2s\,^{3}S$ excited state of $\mathrm{Li}^+$, in the context of determining the lithium nuclear magnetic moment from the hyperfine splitting: the minimization of the variance improved this splitting only marginally below the value already obtained by ordinary energy minimization, while the energy error grew substantially worse. This led them to recommend the conventional variational method over $\sigma^2$ minimization for determining accurate properties \cite{JamesYost1938}.

Stevenson and Crawford, in the helium calculation already cited \cite{Stevenson1938}, noted that minimizing the mean square deviation is ``a more stringent requirement'' than minimizing the energy, and that the resulting optimization ``probably yields a function whose derivatives represent more closely the actual derivatives.''

In a subsequent paper, James and Coolidge \cite{James1939} further formalized the criteria of goodness for approximate wave functions, discussing how symmetry properties and the energy gap to excited states govern the relationship between energy error and wave function error.

Following James and Coolidge, Williamson \cite{Williamson1942} utilized these criteria, including the local energy variance, to assess the accuracy of trial wave functions for the negative hydrogen ion. However, he concluded that $\sigma^2$ is a very poor criterion for overall accuracy, as it changed by only 4\% despite a major reduction in the energy error. The added terms only treated long-range errors; the variance, dominated by the short-range error, was not improved.

\subsection{Discrete-Point Minimization}

Despite these theoretical insights, the impossibility of evaluating $\langle H^2 \rangle$ analytically for many-electron systems stalled progress. Frost achieved a conceptual breakthrough in 1942 \cite{Frost1942}, when he proposed to minimize the variance of the local energy evaluated at a small discrete set of representative points in configuration space. Applied to $\mathrm{H}_2^+$ at the equilibrium distance, this represented the earliest example of a fully deterministic, pre-Monte-Carlo variance-minimization scheme used to generate a wave function without computing any analytical integrals. 
Frost formulated the zero-variance principle in almost exactly its modern form, noting that for an exact solution, ``regardless of what weighting procedure is used the mean square deviation will be zero''. 

In the same paper, he diagnosed the false-convergence problem: a set of points on which the local energy takes a constant but incorrect value, driving the deviation to zero for an incorrect wave function. He stressed that simply increasing the number of points does not guarantee convergence to the correct solution unless the distribution of points is chosen to eventually cover the entire configuration space in the limit.

In a note added in proof he acknowledged, at Pauling's suggestion, that the method was ``closely related'' to Weinstein's 1934 criterion \cite{Weinstein1934}: the link between the lower-bound and least-squares approaches was documented from the beginning, as we have shown, but Frost had arrived at the idea independently before Pauling brought the connection to his attention.

After Frost's 1942 paper, progress stalled for a decade, probably due to the lack of computing power. In 1953, von Mohrenstein \cite{Mohrenstein1953} extended Frost's method: rather than uniformly distributing points and applying a $|\Psi|^2$ weighting factor afterward, he proposed selecting the density of the points to be directly proportional to $|\Psi|^2$: a primitive form of importance sampling. Furthermore, von Mohrenstein identified the need to enforce electron-nucleus and electron-electron cusp conditions to prevent the local energy from diverging at singularities but did not implement them: the required linear correction factor, he noted in a footnote, ``was discovered only later and could not be used'' in his calculations \cite{Mohrenstein1953}. He also observed that unless the singularities are avoided, a few points practically determine the overall error and the values of the parameters.

Bartlett returned to the local energy in 1955, introducing a finite-difference solution of the helium wave equation. As a test of quality he evaluated $E_L$ throughout configuration space and displayed it as a contour map \cite{Bartlett1955}: for the six-term Hylleraas function the local energy is close to its average value ``in only a small region of space,'' diverging to infinity at the singularities of the potential. These represent, to our knowledge, the earliest attempts to map the local energy surface.

Preuss and Trefftz in 1957 optimized a Gaussian trial function for the hydrogen atom using variance minimization instead of energy minimization. They found that it did not generally improve properties such as oscillator strengths, due to the nuclear singularity previously noted by James and Coolidge \cite{Preuss1957}. However, the method did yield a better value at the nucleus, $\Psi(0)$, confirming their prediction that it would be better at describing such short-range properties. 

The following year, Preuss introduced a generalized deviation integral 
\begin{equation}
\delta^2(f) = \int[f(H-E)\Psi]^2\,d\tau,
\end{equation}
weighted by an arbitrary function $f$ of one's own choosing~\cite{Preuss1958}. He was motivated by the observation that standard energy minimization preferentially adjusts the trial function near the nucleus at the expense of other regions relevant to properties like polarization and dipole moments~\cite{Preuss1958}. Demonstrating this method on the hydrogen atom, he showed that setting $f=r$ recovers $\langle r\rangle$ with much better accuracy than standard energy minimization with the same functional form \cite{Preuss1958}.

Preuss returned to the question in 1961, comparing different quality criteria (Temple, Weinstein, Stevenson, and others) and recasting this weighted deviation as the functional
\begin{equation}
\delta^2(f) = \int f^2\,[(H-\langle H\rangle)\Psi]^2\,d\tau,
\end{equation}
built specifically to emphasize whatever region of space governs the expectation value $\langle f\rangle$ one actually wants to estimate \cite{Preuss1961}. Preuss argued that energy minimization is not a sufficient criterion for wave function quality, and showed that standard variance minimization inherently biases the optimization toward regions where $\Psi H \Psi$ is large, potentially degrading the accuracy of other physical observables. On the hydrogen-atom trial function, minimizing $\delta^2(r)$ instead of the energy recovers $\langle r\rangle$ to within 0.7\%, whereas minimizing the energy with the identical functional form is off by 33\%.

With the advent of early digital computers, Frost and coworkers revived the ``local energy method'' in the 1960s. In a paper with Kellogg and Curtis, they applied the method to $\mathrm{H}_2^+$ at the equilibrium distance, focusing on the systematic selection of representative points in configuration space \cite{Frost1960}. With as few as two terms the energy was already within about 0.0006 hartree of the exact value, and adding further terms improved the variance steadily without improving the energy monotonically. By 1961, working with Kellogg, Gimarc, and Scargle \cite{Frost1961}, Frost had replaced the arbitrary point sets of his old method with points and weights chosen by Gauss quadrature: twelve points reproduced the exact $\mathrm{H}_2^+$ energy to six significant figures. Applied to He and H$_2$, the accuracy was more modest (three to four significant figures.)

Gimarc and Frost \cite{Gimarc1963a, Gimarc1963b} extended the method to the lithium atom and to the ground and lowest triplet states of helium, demonstrating its potential but pointing out its limitations: the Gauss quadrature method forced the ``unfortunate compromise'' of using the same exponent for the two lithium shells \cite{Gimarc1963a}, while for excited helium singlet states the variance equations proved ``difficult to solve for the higher roots'' \cite{Gimarc1963b}.
In the final paper of the series, Harriss and Frost \cite{Harriss1964} achieved chemical accuracy for the hydrogen molecule.

Around the same time, Goodisman and Secrest carried out an explicit Weinstein calculation on $\mathrm{H}_2^+$, maximizing the lower bound for a two-parameter trial function. They found that the lower bound was far more sensitive than the energy to the trial function parameters, improving considerably more than the upper bound as flexibility was added \cite{GoodismanSecrest1964}.
They also remarked that varying the trial function to minimize the variance directly ``does not seem to have been performed'', with no reference to Frost's local energy method. It is somewhat ironic that by 1964, parts of this history were already being forgotten.

In a subsequent paper, Goodisman finally discovered Frost's work and gave an alternative derivation of his local energy equations, showing that the local energy method amounts to variance minimization carried out with approximate rather than exact integrals \cite{Goodisman1964}.

\subsection{The Beginning of the Monte Carlo Era}

The foundations of modern VMC were laid by Harold Conroy in a series of twelve papers published in \textit{The Journal of Chemical Physics} between 1964 and 1969 \cite{Conroy1964comm, Conroy1964I, Conroy1964a, Conroy1964b, Conroy1964IV, ConroyBruner1965, ConroyV1967, ConroyBruner1967, Conroy1967, Conroy1967VIII, ConroyMalli1969, Conroy1969X}, beginning with a communication received in October 1963 that already contained all the elements of his research program: the exact cancellation of all electron-nuclei singularities, the Bessel--Clifford correlation factor to cancel the electron-electron singularity, the minimization of the variance with $E_R$ fixed, and the extrapolation to zero variance. These ideas would become the core of Papers I--VII \cite{Conroy1964comm}.

In the first numbered paper of the series, Conroy built one-electron trial functions designed to analytically cancel the Coulomb singularities at the nuclei, ensuring a finite local energy everywhere. He used both the energy and the variance as key diagnostics of how closely the trial functions satisfy the wave equation. He also showed that several excited states of $\mathrm{H}_2^+$ could be optimized directly by using the known excited state energy as the reference energy in the variance functional \cite{Conroy1964I}.

In the next paper, Conroy described his optimization method, introducing Monte Carlo integration to evaluate the matrix elements needed for the linear-variational minimization of $\sigma^2$ \cite{Conroy1964a}. He employed a primitive importance-sampling scheme: he fitted an approximate three-dimensional electron density $D(r)$ built from the nuclear charges, the number of electrons, and a rough estimate of the total energy. He then sampled configurations with a probability proportional to $D(r)$ rather than uniformly.

In the third paper of the series he reported that Monte Carlo residual errors ``do not interfere'' with the mean energy when the wave function is obtained by minimizing $\sigma^2$, but these same errors fed into a RR secular determinant can push its eigenvalues below the true ground state energy \cite{Conroy1964b}.
He also confronted Frost's competing local energy method directly. Frost centers the fit on the current variational estimate $E$, recomputed self-consistently at each iteration; however, since $E$ is not the true eigenvalue $E_0$, Conroy pointed out that the resulting error in the wave function ``will tend to be magnified upon iteration.'' He recommended anchoring the fit instead to any fixed, externally supplied estimate known to lie closer to $E_0$ than $E$ does. 

To extrapolate a final energy from a finite calculation, he introduced ``spoiling plots'': starting from the best available wave function, he continuously varied the coefficient of the single term with the largest overlap with the optimized function away from its optimal value, and extrapolated the resulting $(\sigma,E)$ curve back to $\sigma=0$ using the Temple semicircle as a geometric constraint \cite{Conroy1964b}.

The fourth paper extended the method to two-electron systems by introducing an explicit correlation factor: the Bessel--Clifford $\alpha$ function, designed to cancel the electron-electron singularities. He minimized $\sigma^2$ to optimize the correlation parameters and to rank trial functions of increasing sophistication \cite{Conroy1964IV}. In a short unnumbered communication with Bruner, they adapted the extrapolation scheme to the linear H$_3$ system \cite{ConroyBruner1965}.

The fifth paper generalized the cusp-satisfying construction to an arbitrary number of electrons \cite{ConroyV1967}, extending the excited-state targeting (first demonstrated for one-electron states of $\mathrm{H}_2^+$ in Paper~I) to two-electron systems such as the $^1\Sigma_u^+$ state of $\mathrm{H}_2$.
Conroy's wave function has the modern Slater--Jastrow form, with a correlation function that cancels the interelectronic singularity analytically.

Conroy and Bruner \cite{ConroyBruner1967} provided the first major chemical application of the method, mapping the potential energy surfaces of the three-electron $\mathrm{H}_3$ system and other species like $\mathrm{He}_2^+$ and $\mathrm{He}_2^{++}$.

Simultaneously, the focus shifted toward formalizing the properties of the variance functional itself. Conroy's seventh paper worked out the geometry of the $(\sigma^2,E)$ plane in full, showing that a continuously swept reference energy traces out the edge of the minimum attainable $\sigma^2$ for each energy. From this geometry, he proposed two equivalent extrapolation schemes, whose estimates he reported to be ``invariably a more accurate approximation to the true eigenvalue'' than either the RR upper bound or the Temple lower bound, ``regardless of the quality of the wavefunction'' \cite{Conroy1967}.

The width functional was used yet again by Fraga and Birss \cite{FragaBirss1964}, who formulated a self-consistent field scheme (valid for any state, any multiplicity, with no orthogonality constraints) built on minimizing $\sigma^2$, redefining the Hartree--Fock wave function itself as the uncorrelated function of smallest width.

Other researchers explored the algorithmic challenges of variance minimization. Stanton and Taylor \cite{Stanton1966}, by working out the mathematics of Frost's local energy equations, demonstrated that not every solution of those equations corresponds to a variance minimum: some are maxima, while others are saddle points. Harriss and Roubal \cite{Harriss1968} mapped the variance surfaces for the ground and three lowest excited $\sigma$ states of $\mathrm{HeH}^{2+}$ to find the optimum nonlinear exponents. They found that each surface exhibited a single minimum, but they noted that the optimization was complicated by the extreme sensitivity of the variance to simultaneous parameter changes and the algebraic complexity of the resulting quartic minimization equations.
Harriss and Solev \cite{Harriss1975} studied the convergence of Frost, Kellogg, and Curtis's original iterative solution scheme, showing that particular choices of starting parameters yielded rapid convergence where other choices did not.

Goodisman in 1965 attacked a different problem: bounding an excited-state energy from above using only the eigenvalue of the state immediately below it, again with no orthogonality requirement, demonstrating the method on the excited $^2\Sigma_g^+$ states of $\mathrm{H}_2^+$ \cite{Goodisman1965}.

The technical challenge of evaluating $\langle H^2\rangle$ was directly addressed by Goodisman and Secrest in a 1966 numerical integration study on $\mathrm{H}_2$~\cite{GoodismanSecrest1966}. By introducing a cusp-corrected basis set that analytically removed the $r_{12} \to 0$ singularity, they obtained rigorous Weinstein and Temple lower bounds. However, they warned that minimizing the width to generate excited-state wave functions was untrustworthy: in their calculations the minimization repeatedly returned the ground-state function, or, at larger basis sizes, collapsed into mixtures of excited states.

In a companion study, Goodisman compared energy-minimized with variance-minimized wave functions for the ground state of H$_2$ near its equilibrium distance, finding that variance optimization yielded worse expectation values for almost every property, except $\langle\delta(r_{12})\rangle$ \cite{Goodisman1966}. His conclusion was that ``the only advantage connected with working with the width rather than the energy is the fact that, since one knows the width is zero for the exact wavefunction, one has an idea of how far off one is and how quickly one is improving''. In the history of variance optimization, its role has repeatedly oscillated between a diagnostic tool and an optimization target.

In 1967, Goodisman unified these insights by explicitly identifying Weinstein’s principle as the mathematical foundation underlying both Frost’s and Conroy’s approaches~\cite{Goodisman1967}. Testing it on several states of $\mathrm{H}_2^+$ and $\mathrm{H}_2$, and building on Stanton and Taylor~\cite{Stanton1966}, he ultimately recommended Weinstein calculations as a diagnostic companion to, rather than a replacement for, standard Rayleigh--Ritz calculations.

Lloyd and Delves \cite{Lloyd1969} compared variance minimization with the Rayleigh-Ritz method, focusing on the effects of numerical integration errors. They proved that while RR is inherently superior for the ground state when integrals are exact, variance minimization remains stable at large basis sizes where RR calculations suffer from numerical instability. They suggested using the two methods in conjunction, to estimate the numerical accuracy of the RR energies.

Messmer \cite{Messmer1969}, building on the work of James and Coolidge \cite{James1937}, proposed minimizing the ratio $(\Delta E)^2/\sigma^2$ of the squared energy error to the local energy variance, to maximize the overlap with exact excited state eigenfunctions without requiring explicit orthogonality to the states below it; Choi, Lebeda, and Messmer \cite{Choi1970} then improved the method by removing the requirement that the exact eigenvalue be known in advance by treating it as a variable parameter and establishing exact overlap bounds. 

Cohen and Feldmann shortly thereafter rederived Messmer's functional directly as a generalization of Weinstein's principle, and used this derivation to point out a gap in its justification: nothing guarantees that the optimized wave function's overlap is dominated by the target state rather than spread across lower-lying ones, though they noted this did not appear to be a problem in practice for Messmer's own helium calculations \cite{cohenfeldmann1970b}.

Researchers like Weinhold \cite{weinhold1970} and Cohen and Feldmann \cite{CohenFeldmann1969,cohenfeldmann1970}, during those same years, published extensive generalizations of the Temple, Weinstein and Stevenson--Crawford bounds, exploring how variance constrains other quantum mechanical properties.
Weinhold's review was itself driven by the observation that the literature had become ``widely scattered'' and that ``some needless duplication of effort has resulted'', citing six independent rediscoveries of Eckart's criterion \cite{weinhold1970}: a clear example of a recurrent theme in this history that was already apparent in 1970. 

Yet, despite these mathematical insights, the difficulty of evaluating $\sigma^2$ for many-electron systems meant that these rigorous bounds remained largely a theoretical curiosity. By the end of the decade, variance optimization was already out of fashion, mostly because of the rapid adoption of Gaussian basis sets in electronic structure calculations. 

Within Conroy's own program, the Monte Carlo thread specifically ended even earlier, when Conroy himself abandoned it. The eighth paper of the series replaced stochastic integration with a deterministic, number-theoretic quadrature designed to ``render obsolete\ldots the commonly used Monte Carlo procedure,'' with errors shown to decrease at least as fast as the inverse square of the number of sample points, rather than the inverse square root \cite{Conroy1967VIII}. Applied to four-electron systems in 1969 \cite{ConroyMalli1969}, Conroy described his own earlier Monte Carlo approach as ``now obsolete'' \cite{Conroy1969X}.

\subsection{The Golden Age of Variance Optimization and its Decline}
The development of modern Variational Monte Carlo for many-body systems started in 1965, when McMillan~\cite{McMillan1965} combined the variational principle with the Metropolis algorithm to evaluate the energy of liquid $^4\mathrm{He}$ using a Jastrow trial function.

The history of variance briefly intersected VMC again in 1977 when Coldwell published his ``Zero Monte Carlo Error or Quantum Mechanics is Easier'' \cite{Coldwell1977} paper. Strangely enough, Coldwell's paper cites none of the history retraced here: the constancy of $H\Psi/\Psi$ for an eigenfunction is presented as a fresh observation, with no reference to Frost, Conroy, or any of the literature from the previous fifty years.

It was Umrigar, Wilson, and Wilkins in 1988, however, who gave the method its modern, general-purpose robustness. Their key innovation was to minimize the variance over a large, fixed ensemble of Monte Carlo configurations, reusing the same sample across many optimization steps via correlated sampling rather than redrawing it at every iteration \cite{umrigar1988}. 
Their algorithm was numerically stable for many-parameter Slater-Jastrow wave functions, obtaining smaller statistical errors and improved expectation values compared to the wave functions in common use at the time. It was this specific numerical strategy, not the underlying idea of minimizing $\sigma^2$ (which they credited to Coldwell and to Bartlett's 1955 work, without citing either Frost or Conroy), that sparked the decade-long ``golden age'' of variance minimization during which it became the default optimization strategy in quantum Monte Carlo.
However, the fixed sample technique suffered from numerical problems: one could optimize the parameters to minimize the variance on a fixed ensemble of configurations, only to find that the resulting wave function yielded worse energies and variances when evaluated on a new ensemble.

A note at this point is appropriate: unlike the sparse early history retraced above, from 1988 onward the literature grows rapidly, and we can only follow the papers directly relevant to the choice of the cost function, referring the reader to the standard reviews \cite{Foulkes2001, Austin2012} for the broader development of quantum Monte Carlo methods. 

Alexander, Coldwell, Monkhorst, and Morgan \cite{Alexander1991} in 1991 conducted a systematic comparison of the biased-selection Monte Carlo method against a range of VMC optimization functionals, including the Mean Absolute Deviation (MAD) of the local energy. They showed that the choice of functional significantly impacts the final energy and variance, and that the standard variance functional often overweights regions where the trial function is large.
Using exact values determined via numerical quadrature, they found that Conroy's functional yielded the most accurate eigenvalue in almost every system; in contrast, standard variance minimization achieved the lowest variance but resulted in energies several standard deviations above the optimum \cite{Alexander1991}.

Kent, Needs, and Rajagopal \cite{Kent1999} traced the instabilities of fixed sample variance minimization to the correlated sampling weights: they can grow exponentially when the parameters move far from their sampling values. Variance minimization nonetheless remains the more stable procedure because the positions of the variance minima are robust to finite sampling while those of the energy are not -- an observation essentially identical to Frost's in 1942. They proposed setting the weights to unity, regenerating the ensemble periodically, and limiting outlying local energies, an approach that became standard practice in the following years. Furthermore, they analyzed the distribution of the local energies, discovering that it is not normal, having instead ``fat tails'', an empirical observation that would be rigorously explained by Trail nine years later \cite{Trail2008}.

Snajdr and Rothstein \cite{Snajdr2000} systematically tested the ``commonly held belief'' that variance-optimized functions yield better properties. They studied He, H$_2$, and LiH with up to 32 parameters: the energy-optimized functions ``consistently provide better estimates of non-energy-related properties.'' In their most accurate case, even though the variational energies were statistically indistinguishable, the energy-optimized function yielded errors more than three times smaller, notwithstanding its substantially higher variance.

Their explanation was that, given two wave functions with the same variational energy error, the one with the larger variance has a better guaranteed overlap with the exact eigenfunction. They then invoked the argument (credited to Ceperley as a private communication) that the variance weights the error spectrum by $(E_k-E_0)^2$ and so optimizes ``a higher energy part of the spectrum'', which is James and Coolidge's 1937 result in modern language.

Bressanini, Morosi, and Mella (2002) \cite{Bressanini2002} diagnosed the underlying statistical issue precisely: minimizing the variance (an $L^2$ norm) implicitly assumes that the sampled local energy has a Gaussian distribution, an assumption that is neither theoretically nor practically justified, and a source of the convergence problems that afflict fixed sample variance minimization.
Reframing the problem from a maximum-likelihood point of view, they proposed and tested several alternative functionals on the helium trimer. Among these was the mean absolute deviation from a reference energy, 
\begin{equation}
\int|E_L(\mathbf R)-E_R|\,\Psi(\mathbf R)^2\,d\mathbf R\big/\!\int\Psi(\mathbf R)^2\,d\mathbf R,    
\end{equation}
which is an $L^1$ norm corresponding to a maximum-likelihood estimator for a Laplacian distribution, making it robust to outliers.

They also derived the closed-form local energy distribution for a single-exponential hydrogen atom trial function and found that its tail decays as $E^{-4}$, suggesting that a loss function matched to this exponent should outperform an arbitrary heavy-tailed choice like Cauchy. However, they did not have a general argument for why the local energy distribution should generically have an $E^{-4}$ tail.

Trail in 2008 put these empirical and heuristic observations on rigorous mathematical footing. By classifying four distinct types of local energy singularity, Trail proved that the sampled local energy distribution generically exhibits $E^{-4}$ power-law tails as $|E|\to\infty$, dominated by the nodal surface of the trial wave function, a singularity that survives even when all cusp conditions are satisfied \cite{Trail2008}.

A probability distribution with tail index $\mu$ has a well-defined variance only for $\mu>3$, but a well-defined uncertainty on that variance only for $\mu>5$, since the latter requires a finite fourth moment. 
The sample variance of the local energy is therefore a well-defined quantity to minimize, but the uncertainty on that estimate is formally undefined at any finite sample size. As a direct consequence, the Central Limit Theorem breaks down for the variance estimator itself.

Even as these refinements were being explored, by the 2000s VMC was turning again toward direct energy optimization. Sorella's stochastic reconfiguration, introduced originally to stabilize the fermion sign problem in Green's-function Monte Carlo \cite{Sorella1998}, was recast as a variational algorithm via a generalized Lanczos scheme \cite{Sorella2001}.

Lin, Zhang, and Rappe \cite{LinZhangRappe2000} first applied the Newton method to VMC wave function optimization. It was then improved by Umrigar and Filippi using a reduced-variance Hessian estimator \cite{UmrigarFilippi2005}. Toulouse and Umrigar \cite{ToulouseUmrigar2007} subsequently extended Nightingale and Melik-Alaverdian's linear method \cite{NightingaleMelikAlaverdian2001} to arbitrary nonlinear parameters. Both algorithms prevent ill-conditioning through diagonal regularization.

Because energy minimization was now computationally tractable and was believed to yield superior wave functions overall, variance optimization slowly fell out of fashion once more, often relegated to a preliminary step for generating initial guesses ahead of energy optimization. The motivation was also practical: fixed-node diffusion Monte Carlo (FN-DMC) accuracy is controlled by the nodal surface of the trial function, and energy optimization was found empirically to improve nodes more reliably than variance minimization. Cuzzocrea and coworkers, in a study titled ``The Troubled Story of Variance Minimization,'' documented cases, most often for excited states, but occasionally even for the ground state, where the variance converges nicely while the energy quietly drifts away from the intended eigenstate 
entirely \cite{Cuzzocrea2020}.

Stanton and Taylor's exact solution of the local energy equations had already formally foreseen this mechanism; they showed that fewer minima than basis functions may survive when the basis is not well chosen \cite{Stanton1966}, a precursor to the modern finding that variance landscapes near target eigenstates can dissolve into barrierless plateaus~\cite{Cuzzocrea2020}.

\subsection{The Modern Resurgence of Variance}

Recently, the neural-network VMC community has rediscovered the same heavy-tailed pathology in a different setting. Highly parameterized neural-network ansatzes, like those employed by FermiNet \cite{pfau2020} and PauliNet \cite{Hermann2020}, suffer from extreme statistical noise: Grohs and Nobile proved that for broad, practically relevant ansatz classes, the local energy and gradient estimators are generically heavy-tailed and fail to admit the higher moments needed for reliable convergence; as a result, energy-optimization algorithms can plateau or become unstable unless the estimators are clipped \cite{Grohs2026}.
To overcome this limitation, researchers have integrated robust statistics into energy optimization by clipping local energies. Hermann, Schätzle, and Noé introduced a logarithmically growing clipping function for PauliNet centered at the median \cite{Hermann2020}, a method subsequently adopted in software suites like DeepQMC \cite{deepqmc2023}.
Similarly, FermiNet clips $E_L$ at a fixed multiple of its running standard deviation \cite{pfau2020}, while von Glehn, Spencer, and Pfau use the mean absolute deviation centered at the median (see Appendix~\ref{app:centering}) to determine the clipping window \cite{vonglehn2023}.

Fu et al. \cite{Fu2024} utilized the empirical linear relationship between energy and variance to perform variance extrapolation in NN-VMC, masking local energies beyond three standard deviations of the median before fitting.

Furthermore, state-of-the-art scaling of NQS to millions of parameters via minimum-step stochastic reconfiguration (MinSR) \cite{Chen2024} relies on this zero-variance extrapolation principle \eqref{eq:extrapolation} to estimate exact ground state energies, drawing upon a geometric boundary first mapped in Conroy's 1964 spoiling plots and formalized in his 1967 analysis \cite{Conroy1964b,Conroy1967}.

More recently, Misery et al. \cite{Misery2026} showed that sampling $|\Psi|^2$ yields high-variance gradient estimators, introducing ``overdispersed sampling'' to flatten the distribution and stabilize training. Others have instead targeted the objective directly: one proposal minimizes a logarithmically compressed variance of the local energy \cite{Jin2026}.

In a return to variance's original role as a diagnostic rather than a target, Shi, Cui, and Zhou revisited the local energy variance as a convergence criterion for neural-network wave functions \cite{Shi2025}.

What is absent from the literature is a systematic comparison of variance minimization against its more outlier-robust statistical alternatives. This is one of the purposes of the present paper: alongside energy, variance, and MAD, we test a maximum-likelihood loss whose tail is analytically matched to Trail's derived exponent, together with the generic Cauchy-type loss from the same family, using the century-old Guillemin--Zener wave function for $\mathrm{H}_2^+$ \cite{GuilleminZener1929}.

\section{Theoretical Background}\label{sec:theory}

In this section, we derive the approximate relationship between energy error and variance. We then define the cost functionals built from a statistical maximum-likelihood perspective, demonstrating why robust $L^1$ and logarithmic norms provide a superior alternative to the standard $L^2$ variance in the presence of heavy-tailed local energy distributions.
To fix the notation, we consider a normalized trial wave function $|\Psi\rangle$ expanded in the exact eigenbasis of the Hamiltonian $H$,
\begin{flalign}
&|\Psi\rangle = \sum_n c_n|\Psi_n\rangle, \quad \sum_n |c_n|^2 = 1, && \notag \\
&\langle H\rangle = \sum_n |c_n|^2 E_n, && \notag \\
&\sigma^2 = \sum_n |c_n|^2(E_n-\langle H\rangle)^2 \;\geq 0 &&
\end{flalign}
\subsection{The Energy-Variance Relationship}

In modern NQS literature, a linear relationship between the variational energy and the variance is observed as the neural network parameters are increased. Fu et al.~\cite{Fu2024} noted that ``the fundamental reason for the linear relationship is not fully clear'', although it follows directly from expanding the wave function error, as first formulated by James and Coolidge \cite{James1937}.

Let us assume a normalized trial wave function $|\Psi\rangle$ that approximates the exact ground state $|\Psi_0\rangle$ with exact energy $E_0$. We can expand the trial function as a mixture of the exact ground
state and an orthogonal, normalized error function $|\Psi_x\rangle$:
\begin{equation}
  |\Psi\rangle = \sqrt{1 - \epsilon^2} |\Psi_0\rangle + \epsilon |\Psi_x\rangle,
  \label{eq:psi_expansion}
\end{equation}
where $\epsilon^2 = 1 - S^2$ represents the total weight of the error in the wave function, $\langle \Psi_0 | \Psi \rangle = S$ and
 $\langle \Psi_0 | \Psi_x \rangle = 0$.

The variational energy, $E = \langle \Psi | H | \Psi \rangle$, evaluates to
\begin{equation}
  E = (1 - \epsilon^2)E_0 + \epsilon^2 \langle \Psi_x | H | \Psi_x \rangle.
  \label{eq:e_variational}
\end{equation}
The error in the energy, $\Delta E = E - E_0$, is therefore proportional to the square of the wave function error:
\begin{equation}
  \Delta E = \epsilon^2 \langle \Psi_x | H - E_0 | \Psi_x \rangle.
  \label{eq:delta_e}
\end{equation}

Next, we evaluate the variance $\sigma^2 = \langle \Psi | H^2 | \Psi \rangle
- E^2$. Expanding the expectation value of $H^2$ yields
\begin{equation}
  \langle H^2 \rangle = (1 - \epsilon^2)E_0^2 + \epsilon^2 \langle \Psi_x | H^2 | \Psi_x \rangle.
\end{equation}
Substituting this and $E = E_0 + \Delta E$ into the variance expression, we obtain
\begin{equation}
  \sigma^2 = \epsilon^2 \langle \Psi_x | (H - E_0)^2 | \Psi_x \rangle - \Delta E^2.
\end{equation}
Because $\Delta E$ scales as $\mathcal{O}(\epsilon^2)$, the $\Delta E^2$ term scales as $\mathcal{O}(\epsilon^4)$ and can be neglected in the limit of a
good wave function ($\epsilon \to 0$). The variance thus simplifies to the leading order:
\begin{equation}
  \sigma^2 \approx \epsilon^2 \langle \Psi_x | (H - E_0)^2 | \Psi_x \rangle.
  \label{eq:variance_approx}
\end{equation}
Both $\Delta E$ and $\sigma^2$ scale as $\epsilon^2$, so by taking their ratio we establish a linear proportionality:
\begin{equation}
  \Delta E_{\mathrm{eff}} = \frac{\sigma^2}{\Delta E} \approx
  \frac{\langle \Psi_x | (H - E_0)^2 | \Psi_x \rangle}
  {\langle \Psi_x | H - E_0 | \Psi_x \rangle}.
  \label{eq:delta_e_eff}
\end{equation}
Here, $\Delta E_{\mathrm{eff}}$ corresponds to the spectral-weighted average excitation energy of the error state $|\Psi_x\rangle$.
James and Coolidge \cite{James1937} employed an equivalent expansion to relate $\Delta E$ to the wave function error $q^2$: their Eq.~(19), $q^2 \approx (\Delta E)^2 K^2/\sigma^2$, gathers wave function error, energy
error, and variance into a single relation. Messmer generalized the same relation to excited states thirty years later \cite{Messmer1969}.

This relationship holds for a given wave function. Consider now the case when a trial wave function is systematically improved, for instance by expanding the basis set or using more correlation factors or increasing the size of a neural network. Expanding the error function $|\Psi_x\rangle$ in the exact eigenstates of the Hamiltonian, the energy error and the variance become:
\begin{align}
  \Delta E &= \sum_{n>0} c_n^2 (E_n - E_0), \label{eq:e_expansion} \\
  \sigma^2 &\approx \sum_{n>0} c_n^2 (E_n - E_0)^2, \label{eq:var_expansion}
\end{align}
where $c_n = \epsilon \langle \Psi_n | \Psi_x \rangle$ are the error coefficients.

If the optimization process acts to uniformly suppress the error spectrum, this mathematically implies that the error coefficients scale according to a global factor $\lambda$ while maintaining a roughly constant relative distribution $p_n$:
\begin{equation}
  c_n \approx \lambda p_n.
\end{equation}
Under this uniform scaling hypothesis, the global scaling factor $\lambda^2$ factors out of both Eq.~\eqref{eq:e_expansion} and Eq.~\eqref{eq:var_expansion}. Consequently, their ratio $\Delta E_{\mathrm{eff}}$ becomes independent of $\lambda$ and acts as a strict constant:
\begin{equation}
  \Delta E_{\mathrm{eff}} \approx
  \frac{\sum_{n>0} p_n^2 (E_n - E_0)^2}{\sum_{n>0} p_n^2 (E_n - E_0)}
  = \text{const.}
\end{equation}
Uniform scaling is sufficient but not necessary: Fu et al.~\cite{Fu2024} derive the same linearity by assuming that the error concentrates near a single excitation energy $E'$, while Chen and Heyl \cite{Chen2024} employ what is essentially the uniform scaling hypothesis.

Rearranging Eq.~\eqref{eq:delta_e_eff} yields a linear relationship between the energy and the variance:
\begin{equation}
  E \approx E_0 + \left( \frac{1}{\Delta E_{\mathrm{eff}}} \right) \sigma^2.
  \label{eq:extrapolation}
\end{equation}
Thus, a plot of $E$ against $\sigma^2$ for a sequence of systematically improved wave functions will yield a straight line whose $y$ intercept is an estimate of the true ground state energy $E_0$, but neither an upper nor a lower bound.

This derivation also shows when this linearity fails, such as when the addition of variational parameters disproportionately cures a specific high-energy error scale (such as a correction only to a nuclear cusp). In this scenario, the relative error distribution $p_n$ changes. This alters $\Delta E_{\mathrm{eff}}$, producing non-linear curvature or inflection points in the curve.

Conroy \cite{Conroy1967} analyzed the geometry of the $(\sigma^2,E)$ plane, proving that for an incomplete basis set the allowed values of $\sigma^2$ and $E$ fall within a bounded region delimited by the Weinstein and Temple formulas. Because the origin $(0, E_0)$ is generally excluded unless the basis is complete, energy and variance optimization exhibit an intrinsic trade-off: minimizing variance along the boundary forces an increase in the energy, while driving $E$ toward its variational minimum increases
 $\sigma^2$.
 
How the relation of Eq.~\eqref{eq:extrapolation} reached the modern literature is again an example of forgotten history. Kwon, Ceperley, and Martin observed the linearity empirically in the electron gas, stressing that there is ``no fundamental reason'' for it to hold \cite{Kwon1998}. Independently, Sorella rediscovered the same extrapolation to benchmark his Lanczos-improved variational energies \cite{Sorella2001}. Kashima and Imada attribute the underlying argument to Sorella's generalized Lanczos paper \cite{Kashima2001}. From there the relation reached the NN-VMC
extrapolations of Fu et al.\ and Chen and Heyl, who cite Kwon et al.\ and Kashima and Imada \cite{Fu2024,Chen2024}. None of these modern papers seem to be aware of the work that started in 1937 and continued for at least thirty years.

\subsection{Cost Functionals and Statistical Robustness}
\label{sec:objfunc}

In VMC, integrals are evaluated as statistical averages over a set of $N$ configurations (walkers) $\mathbf{R}_i$ sampled from a probability distribution, typically $|\Psi(\mathbf{R})|^2$. The mathematical form of the cost functional reflects the underlying statistical model assumed for local energy fluctuations.
We compare five cost functionals:
\begin{flalign*}
&\quad \mathcal L_E = \big\langle E_L\big\rangle \\
&\quad \mathcal L_{\mathrm{Var}} = \big\langle\,(E_L - E_R)^2\,\big\rangle \\
&\quad \mathcal L_{\mathrm{MAD}} = \big\langle\,|E_L - E_R|\,\big\rangle \\
&\quad \mathcal L_{\mathrm{C}} = \big\langle\,\log\!\big(1+(E_L-E_R)^2/s^2\big)\,\big\rangle \\
&\quad \mathcal L_{-4} = \big\langle\,\log\!\big(1+(E_L-E_R)^4/s^4\big)\,\big\rangle
\end{flalign*}
where $s$ is an adjustable scale parameter. Four of these objectives are written as functions of a reference energy $E_R$, which is not necessarily forced to be $\langle E_L\rangle$: unlike the energy, which has no such freedom, $\mathcal L_{\mathrm{Var}}$, $\mathcal L_{\mathrm{MAD}}$, $\mathcal L_{\mathrm C}$, and $\mathcal L_\mathrm{-4}$ each admit more than one choice of $E_R$. The choice of $E_R$ directly influences both the statistical robustness of the optimization and the computational cost of evaluating exact gradients (see Appendix~\ref{app:centering}).

Note that $\mathcal L_{\mathrm{Var}}$ coincides with the variance only for $E_R = \langle E_L \rangle$: in general, it is the second moment about $E_R$, $\langle (E_L-E_R)^2 \rangle = \sigma^2 + (\langle E_L \rangle -
E_R)^2$, and the two objectives differ whenever $E_R$ is held fixed. Trail criticized the inconsistent usage of ``variance'' in the literature \cite{Trail2008}.

Following the maximum-likelihood framework introduced by Bressanini, Morosi, and Mella \cite{Bressanini2002}, minimizing $\mathcal{L}_{\mathrm{Var}}$ implicitly assumes normally distributed local energy fluctuations, for which the optimal location parameter $E_R$ is the sample mean.
However, as rigorously proved by Trail \cite{Trail2008}, local energy distributions of Coulomb systems generically exhibit $E^{-4}$ power-law tails. Consequently, the fourth moment diverges, so the variance of the variance estimator is infinite, rendering the $L^2$ cost functional formally ill-conditioned.

To mitigate this instability, one can construct robust estimators by adopting likelihood functions with heavier tails. Replacing the Gaussian distribution with an exponential (Laplace) distribution yields the Mean Absolute Deviation functional, $\mathcal{L}_{\mathrm{MAD}}$. Minimizing $\mathcal{L}_{\mathrm{MAD}}$ requires setting $E_R$ to the sample median. Unlike the mean, the median remains insensitive to extreme configurations in asymmetric or heavy-tailed distributions.

The Cauchy ($\mathcal L_{\mathrm{C}}$) and $\mathcal L_{-4}$ cost functionals model even heavier tail behaviors. Both represent negative
log-likelihoods of a generalized Student-$t$-type distribution $(1+|E_L-E_R|^\mu/s^\mu)^{-1}$ with $\mu=2$ and $\mu=4$, respectively, differing primarily in how strongly they penalize extreme outliers. While the Cauchy loss ($\mu=2$) provides an ad hoc heavy-tailed alternative chosen purely for empirical robustness, $\mathcal L_{-4}$ ($\mu=4$) directly incorporates the asymptotic exponent derived analytically by Trail for QMC local energy distributions.

Standard VMC implementations usually set $E_R$ to the running arithmetic mean for algorithmic convenience. While this is the correct choice for $\mathcal{L}_{\mathrm{Var}}$, skewness causes the mean to introduce the exact bias these robust estimators are designed to suppress. In this work, we use the sample mean for $\mathcal{L}_{\mathrm{Var}}$. For $\mathcal{L}_{\mathrm{MAD}}$, we set $E_R$ to the sample median. For $\mathcal{L}_{\mathrm{C}}$ and $\mathcal{L}_{-4}$, the exact likelihood implies a self-consistent weighted mean; to avoid solving a nonlinear equation at every optimization step, we employed the sample median as well. This preserves the outlier-rejection characteristics of the estimators while keeping gradient evaluations exact, as detailed in Appendix~\ref{app:centering}.

\subsection{An Alternative Look at Optimization Functionals: the Trial Potential}

These cost functionals admit an equivalent physical interpretation. For any $\Psi$ and constant $E_R$, inverting the Schr\"odinger equation defines an effective trial potential
\[
V_T(\mathbf R) \;\equiv\; E_R - \frac{\hat T \Psi(\mathbf R)}{\Psi(\mathbf R)} ,
\]
for which $\Psi$ is an exact eigenstate; since $E_L=\hat T\Psi(\mathbf R)/\Psi(\mathbf R)+V(\mathbf R)$, this simply gives $V(\mathbf R)-V_T(\mathbf R)=E_L-E_R$ for every configuration and every $E_R$. Choosing a metric $d(\cdot)$ and minimizing a cost functional $d(E_L-E_R)$ is therefore equivalent to minimizing $d(V_T(\mathbf R)-V(\mathbf R))$ under the same metric: optimization fits the trial potential to the real one. Consequently, for example, minimizing the variance of the local energy is equivalent to choosing the $L^2$ norm to measure distances.

The idea is not new. In 1961 Preuss showed by inverting the Schr\"odinger equation that variance optimization can be interpreted as the minimization of an ``error potential'': the potential that must be added to $H$ for $\Psi$ to become an exact eigenstate, providing a rough measure of its deviation from the exact wave function \cite{Preuss1961}. He credited the inversion to Laforgue \cite{Laforgue1954}, and reused the concept in his later work \cite{Preuss1964i, Preuss1964}.

Similarly, Turbiner and Guevara \cite{turbinerguevara2007} inverted the Schr\"odinger equation to construct functional forms for electronic correlation factors.

\section{Computational Methodology}\label{sec:methods}

To evaluate the influence of alternative cost functionals on trial wave function quality, we performed VMC calculations on the $\mathrm{H}_2^+$ molecular ion across its potential energy curve. Because exact Born--Oppenheimer energies for $\mathrm{H}_2^+$ are available to high accuracy \cite{batesledsham1953}, absolute energy errors provide an essentially exact benchmark. We tested two wave functions of increasing flexibility:

\begin{description}

  \item[\textbf{Simple Molecular Orbital (MO)}] The symmetrized combination of two Slater-type orbitals with a single variational exponent $a$,
  \[
  \Psi_{\mathrm{MO}} = e^{-a r_A} + e^{-a r_B},
  \]
where $r_A$ and $r_B$ denote the electron-nucleus distances. The electron-nuclear cusp is not enforced, ensuring that the local energy distribution exhibits the characteristic heavy tails described in Sec.~\ref{sec:theory}.

  \item[\textbf{Guillemin--Zener (GZ)}] A flexible, two-parameter wave function written in prolate spheroidal coordinates $\lambda=(r_A+r_B)/R$ and $\mu=(r_A-r_B)/R$, where $R$ is the internuclear separation. The standard two-parameter Guillemin--Zener function \cite{GuilleminZener1929} for the $1s\sigma_g$ ground state takes the form
\[
\Psi_{\mathrm{GZ}} = \exp\!\big(-\tfrac12 a R\lambda\big)\,
\cosh\!\big(\tfrac12 b R\mu\big),
\]
  where $a$ and $b$ are two nonlinear variational parameters and setting $a=b$ recovers the MO form above.
\end{description}

The MO ansatz was optimized with three cost functionals ($\mathcal{L}_E$, $\sigma^2$, and MAD), while the GZ ansatz was optimized with all five, at each internuclear distance from $R=0.2\,a_0$ to $10.0\,a_0$ in steps of $0.2\,a_0$.

All optimizations used 1000 walkers. The energy was minimized with the stochastic reconfiguration algorithm \cite{Sorella2001}; the remaining functionals were optimized with a standard fixed-ensemble algorithm \cite{umrigar1988}. All optimizers were iterated to convergence. Optimizations proceeded sequentially along the curve, each started from the parameters converged at the preceding internuclear distance, with the same protocol for every cost functional.

For the GZ ansatz, $\mathcal{L}_{\mathrm{C}}$ and $\mathcal{L}_{-4}$ were additionally optimized at five scales, $s = 0.10$, $0.05$, $0.03$, $0.01$, and $0.005$ hartree, over the same $R$ grid.

After optimization, each wave function was re-evaluated in an independent VMC simulation of 1000 walkers propagated for 500 blocks of 1000 steps each; the absolute energy error is the difference between the final energy estimate of this run and the exact value \cite{batesledsham1953}.

The reference energy $E_R$ follows Sec.~\ref{sec:objfunc}: the sample mean for $\mathcal{L}_{\mathrm{Var}}$, the sample median for all other functionals, with gradients evaluated as described in Appendix~\ref{app:centering}.

Statistical uncertainties, estimated from block averages, are much smaller than the symbol sizes in the reported error plots.

\section{Results and Discussion}\label{sec:results}

\subsection{Simple Molecular Orbital Ansatz}

\begin{figure}[htbp]
  \centering
  \includegraphics[width=\columnwidth]{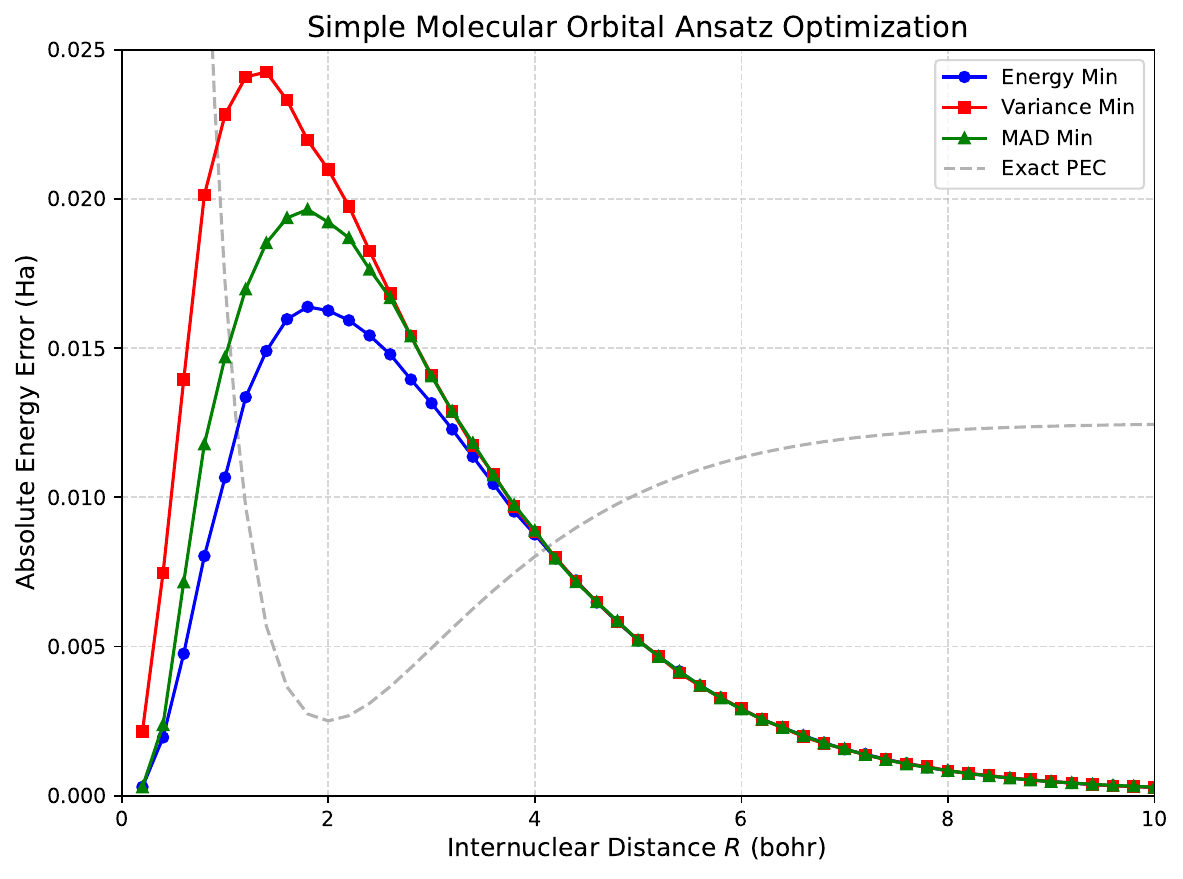}
  \caption{Absolute energy errors for $\mathrm{H}_2^+$ as a function of
  internuclear distance $R$, comparing Energy, Variance, and MAD
  optimization using a simple MO ansatz. The exact potential energy curve is superimposed as a dashed gray line in the background (unmarked secondary axis)
  to highlight the repulsive, equilibrium, and dissociation regions.}
  \label{fig:mo_errors}
\end{figure}

The absolute energy errors relative to the exact numerical solution of $\mathrm{H}_2^+$ for the three cost functionals ($\mathcal{L}_E$, $\sigma^2$ and MAD) are plotted in Figure \ref{fig:mo_errors}. The energy-optimized wave function, by construction, yields the lowest variational energy (within statistical uncertainty).
MAD optimization systematically outperforms variance optimization, especially at small $R$.

This empirically demonstrates the bias inherent in $L^2$ variance minimization: the quadratic penalty heavily weights outlier configurations near the Coulomb singularities, causing the optimizer to converge to a suboptimal region of parameter space. The $L^1$ MAD cost functional, being robust to these heavy-tailed outliers, finds a better wave function.

For $R > 4.0\,a_0$, the performance of all three cost functionals converges. As $R\to\infty$ the system separates into an isolated H atom plus a non-interacting proton, while as $R\to0$ the system approaches the $\mathrm{He}^+$ united atom. In both cases the exact ground state is a single exponential, so the one-parameter ansatz becomes exact and the error goes to zero regardless of the cost functional.

\subsection{Guillemin--Zener Ansatz}

\begin{figure}[htbp]
  \centering
  \includegraphics[width=\columnwidth]{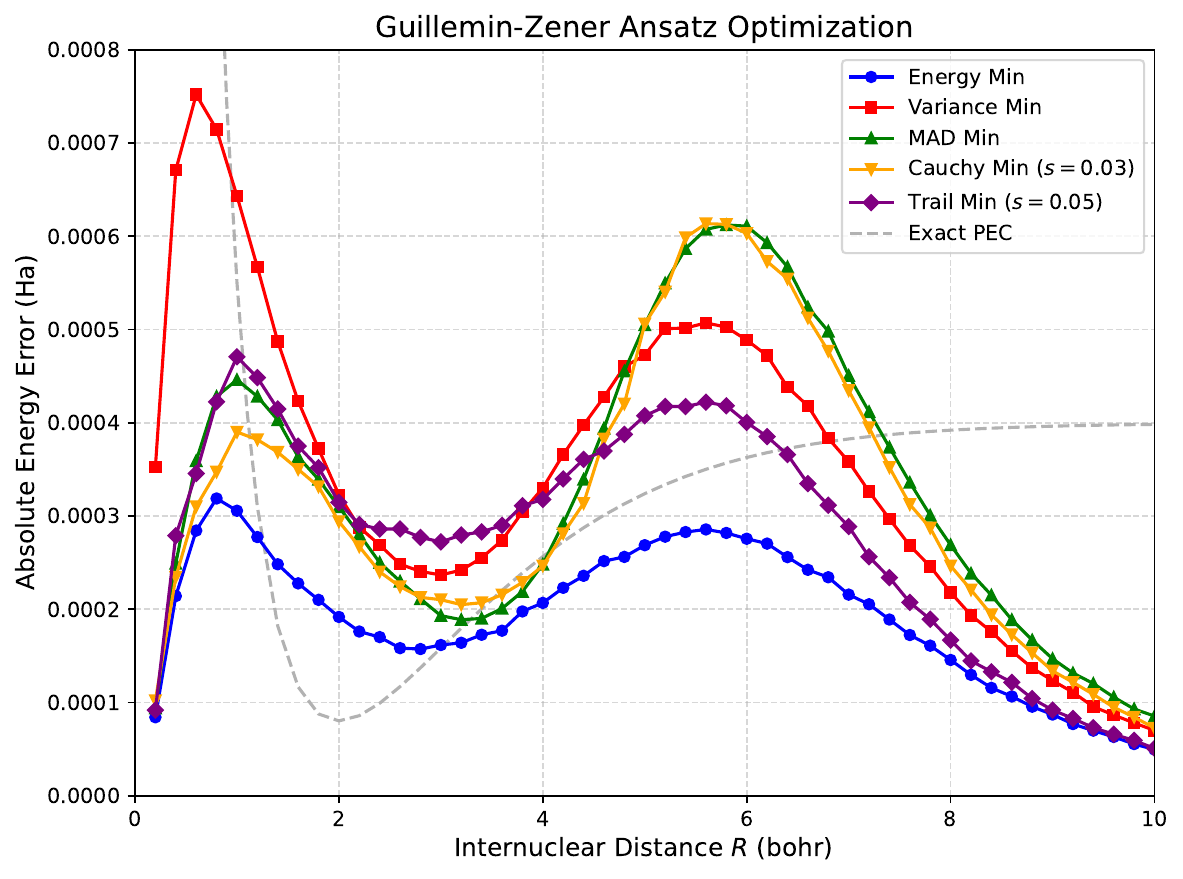}
  \caption{Absolute energy errors for $\mathrm{H}_2^+$ using the
  Guillemin--Zener ansatz minimized with Energy, Variance, MAD,
  $\mathcal L_{\mathrm{C}}$ and $\mathcal L_{-4}$ cost functions. The exact potential energy curve is superimposed as a dashed gray line in
  the background (unmarked secondary axis) to guide the
  eye.}
  \label{fig:gz_errors}
\end{figure}

The flexible Guillemin--Zener ansatz achieves sub-chemical accuracy ($< 1\,\mathrm{kcal/mol} \approx 1.6 \times 10^{-3}$ hartree) across the entire potential energy curve, regardless of the cost functional. The absolute error profiles for the GZ wave function optimized with the five functionals are shown in Figure \ref{fig:gz_errors}, superimposed over the exact potential energy curve (dashed gray line).
All cost functionals yield a characteristic double-peaked error profile, with local maxima at $R \approx 0.6$--$1.2\,a_0$ and at $R\approx 5.6$--$6.0\,a_0$, separated by a distinct local minimum near $R \approx 2.8$--$3.2\,a_0$, beyond the equilibrium bond distance $R_{\mathrm{eq}} \approx 2.0\,a_0$.
This reflects the specific GZ functional form, which is well suited to describe the bonding region but is slightly less flexible in the repulsive region of the potential energy curve and in the intermediate dissociation region, where atomic polarization cannot be fully captured by two exponential parameters.

The comparative behavior of the cost functionals shows that, as expected, energy minimization consistently yields the lowest error. MAD minimization outperforms variance minimization in the bonding and intermediate regions, up to $R \simeq 4.8\,a_0$; beyond this point, variance optimization gives the better wave function of the two. 

At large $R$, the $\mathrm{H}_2^+$ ground state approaches the symmetric combination of hydrogenic $1s$ orbitals on the two protons, so the electron is effectively always near one of them. The local energy distribution becomes less skewed and there is effectively only one Coulomb singularity which the GZ form
describes well. In this regime the $L^2$ variance cost functional is no longer ill-conditioned, and its quadratic penalty becomes an advantage over MAD.

In addition to the $L^2$ and $L^1$ norms, we tested the $\mathcal L_{\mathrm{C}}$ and the $\mathcal L_{-4}$ functionals. For both we found that the optimization is highly sensitive to the choice of the scale parameter $s$. The Cauchy functional shows a crossover behavior: for small $s$, the logarithmic penalty saturates rapidly, causing the functional to behave like the $L^1$ MAD, whereas, for large $s$ the functional mimics the $L^2$ variance. This is explained by considering the loss $\ell(x) = \log(1 + x^2/s^2)$, where $x = E_L - E_R$. If $s \gg |x|$, then $\log(1 + x^2/s^2)\approx x^2/s^2$ and the Cauchy cost functional reduces to a multiple of the variance. On the other hand, if $x^2/s^2 \gg 1$, $\ell(x) \approx 2\log(|x|/s)$: the cost function grows only logarithmically, heavily suppressing outliers and behaving somewhat like MAD.

At the intermediate scale parameter $s=0.03$, shown in Figure \ref{fig:gz_errors}, the Cauchy functional achieves a slightly lower error than MAD at the equilibrium distance. However, the gain is marginal and far from the accuracy of energy minimization. Given that $\mathcal{L}_{\mathrm{C}}$ requires a manual scan to optimize the scale parameter $s$, it offers no practical advantage over the parameter-free MAD in the bonding region, at least for the system studied here.

Focusing on $\mathcal L_{-4}$, the choice $s=0.05$ as shown in Figure \ref{fig:gz_errors} (diamonds) performs well. Up to the equilibrium region it achieves an accuracy comparable to MAD. However, beyond $R \approx 4.8\,a_0$, this cost functional systematically outperforms both MAD and $\sigma^2$, providing the lowest absolute error of all the non-energy functionals as the system approaches the separated-atom limit.

At $R=10.0\,a_0$ it is statistically indistinguishable from the energy-optimized wave function, and is significantly better than both the variance and MAD.

This indicates that matching the cost functional to the true underlying heavy-tailed distribution yields a real improvement in wave function quality, particularly in the asymptotic tail of the potential energy curve. This, however, requires a careful tuning of $s$.

For the GZ ansatz, variance minimization is never the best choice at any $R$: MAD is better in the bonding region, while $\mathcal{L}_{-4}$ is better in the dissociation region.

We compared wave functions by their absolute variational energy error $|\langle H\rangle-E_0|$, the metric most relevant to VMC practice, but it is worth recalling again that a wave function optimized for a lower energy is not necessarily optimized for a lower variance, a better overlap with the exact state, or more accurate expectation values of other observables.

\section{Conclusions}\label{sec:conclusions}

In this work, we retraced the history of variance-based wave function optimization from 1928 to the present, recovering results rarely cited in the modern literature, such as the energy-variance relationship, that are relevant for modern NQS and NN-VMC approaches. We then tested five cost functionals on $\mathrm{H}_2^+$ to show, on a system with an essentially exact solution, that the $L^2$ variance is never the best of the available alternatives. The parameter-free Mean Absolute Deviation acts as a reliable cost functional in the bonding region by rejecting extreme outliers, whereas the $\mathcal{L}_{-4}$ functional, which incorporates the $E^{-4}$ asymptotic tail behavior, systematically outperforms both MAD and variance in the dissociation regime.

The history reviewed here also suggests possible directions that earlier work could only sketch on few-electron systems but that modern NQS architectures can explore at scale. For example, Conroy's spoiling plots could be tested within a single trained NQS (via dropout or weight noise) to map the $(\sigma^2, E)$ boundary at a fraction of the cost of training a family of networks. Similarly, Preuss's idea of weighting the variance by a target operator $f$ to selectively improve a specific observable $\langle f \rangle$ suggests a multi-objective extension of NQS training: including force or dipole operators directly in the loss function to produce wave functions tailored for specific properties.

In the end, the central lesson of the reviewed history is that the choice of the cost functional is not a minor algorithmic detail but a statistical assumption about the local energy distribution, and that exploring alternative cost functionals and different weighting schemes can yield genuine improvements.

\medskip
\noindent\textit{Author's note.} This manuscript is being posted exclusively to arXiv, at least in its present form, because papers of this kind, featuring a very long historical section and a small and deliberately simple test system, in my experience, do not fit well in the current editorial landscape of chemistry and physics journals.

The long historical section and the small test system are both explicit choices: the first, because recovering forgotten work is one of the primary purposes of this paper; the second, because $\mathrm{H}_2^+$ is one of the few systems with a known essentially exact solution to compare against.

In this work, I made every effort to track down and cite every paper directly related to wave function variance optimization, at least up to the early 1970s. I have not tried, however, to track every modern NN-VMC paper that uses variance or related statistical devices.

Comments and corrections are genuinely welcome.

\bibliography{referenceswithnames_corrected_annotated}

\begin{thebibliography}{91}%
\makeatletter
\providecommand \@ifxundefined [1]{%
 \@ifx{#1\undefined}
}%
\providecommand \@ifnum [1]{%
 \ifnum #1\expandafter \@firstoftwo
 \else \expandafter \@secondoftwo
 \fi
}%
\providecommand \@ifx [1]{%
 \ifx #1\expandafter \@firstoftwo
 \else \expandafter \@secondoftwo
 \fi
}%
\providecommand \natexlab [1]{#1}%
\providecommand \enquote  [1]{``#1''}%
\providecommand \bibnamefont  [1]{#1}%
\providecommand \bibfnamefont [1]{#1}%
\providecommand \citenamefont [1]{#1}%
\providecommand \href@noop [0]{\@secondoftwo}%
\providecommand \href [0]{\begingroup \@sanitize@url \@href}%
\providecommand \@href[1]{\@@startlink{#1}\@@href}%
\providecommand \@@href[1]{\endgroup#1\@@endlink}%
\providecommand \@sanitize@url [0]{\catcode `\\12\catcode `\$12\catcode `\&12\catcode `\#12\catcode `\^12\catcode `\_12\catcode `\%12\relax}%
\providecommand \@@startlink[1]{}%
\providecommand \@@endlink[0]{}%
\providecommand \url  [0]{\begingroup\@sanitize@url \@url }%
\providecommand \@url [1]{\endgroup\@href {#1}{\urlprefix }}%
\providecommand \urlprefix  [0]{URL }%
\providecommand \Eprint [0]{\href }%
\providecommand \doibase [0]{https://doi.org/}%
\providecommand \selectlanguage [0]{\@gobble}%
\providecommand \bibinfo  [0]{\@secondoftwo}%
\providecommand \bibfield  [0]{\@secondoftwo}%
\providecommand \translation [1]{[#1]}%
\providecommand \BibitemOpen [0]{}%
\providecommand \bibitemStop [0]{}%
\providecommand \bibitemNoStop [0]{.\EOS\space}%
\providecommand \EOS [0]{\spacefactor3000\relax}%
\providecommand \BibitemShut  [1]{\csname bibitem#1\endcsname}%
\let\auto@bib@innerbib\@empty
\bibitem [{\citenamefont {Umrigar}\ \emph {et~al.}(1988)\citenamefont {Umrigar}, \citenamefont {Wilson},\ and\ \citenamefont {Wilkins}}]{umrigar1988}%
  \BibitemOpen
  \bibfield  {author} {\bibinfo {author} {\bibfnamefont {C.~J.}\ \bibnamefont {Umrigar}}, \bibinfo {author} {\bibfnamefont {K.~G.}\ \bibnamefont {Wilson}},\ and\ \bibinfo {author} {\bibfnamefont {J.~W.}\ \bibnamefont {Wilkins}},\ }\bibfield  {title} {\bibinfo {title} {{Optimized trial wave functions for quantum Monte Carlo calculations}},\ }\href@noop {} {\bibfield  {journal} {\bibinfo  {journal} {Phys. Rev. Lett.}\ }\textbf {\bibinfo {volume} {60}},\ \bibinfo {pages} {1719} (\bibinfo {year} {1988})}\BibitemShut {NoStop}%
\bibitem [{\citenamefont {Sorella}(2001)}]{Sorella2001}%
  \BibitemOpen
  \bibfield  {author} {\bibinfo {author} {\bibfnamefont {S.}~\bibnamefont {Sorella}},\ }\bibfield  {title} {\bibinfo {title} {{Generalized Lanczos algorithm for variational quantum Monte Carlo}},\ }\href@noop {} {\bibfield  {journal} {\bibinfo  {journal} {Phys. Rev. B}\ }\textbf {\bibinfo {volume} {64}},\ \bibinfo {pages} {024512} (\bibinfo {year} {2001})}\BibitemShut {NoStop}%
\bibitem [{\citenamefont {Nightingale}\ and\ \citenamefont {Melik-Alaverdian}(2001)}]{NightingaleMelikAlaverdian2001}%
  \BibitemOpen
  \bibfield  {author} {\bibinfo {author} {\bibfnamefont {M.~P.}\ \bibnamefont {Nightingale}}\ and\ \bibinfo {author} {\bibfnamefont {V.}~\bibnamefont {Melik-Alaverdian}},\ }\bibfield  {title} {\bibinfo {title} {{Optimization of Ground- and Excited-State Wave Functions and van der Waals Clusters}},\ }\href@noop {} {\bibfield  {journal} {\bibinfo  {journal} {Phys. Rev. Lett.}\ }\textbf {\bibinfo {volume} {87}},\ \bibinfo {pages} {043401} (\bibinfo {year} {2001})}\BibitemShut {NoStop}%
\bibitem [{\citenamefont {Umrigar}\ \emph {et~al.}(2007)\citenamefont {Umrigar}, \citenamefont {Toulouse}, \citenamefont {Filippi}, \citenamefont {Sorella},\ and\ \citenamefont {Hennig}}]{Umrigar2007}%
  \BibitemOpen
  \bibfield  {author} {\bibinfo {author} {\bibfnamefont {C.~J.}\ \bibnamefont {Umrigar}}, \bibinfo {author} {\bibfnamefont {J.}~\bibnamefont {Toulouse}}, \bibinfo {author} {\bibfnamefont {C.}~\bibnamefont {Filippi}}, \bibinfo {author} {\bibfnamefont {S.}~\bibnamefont {Sorella}},\ and\ \bibinfo {author} {\bibfnamefont {R.~G.}\ \bibnamefont {Hennig}},\ }\bibfield  {title} {\bibinfo {title} {{Alleviation of the Fermion-Sign Problem by Optimization of Many-Body Wave Functions}},\ }\href@noop {} {\bibfield  {journal} {\bibinfo  {journal} {Phys. Rev. Lett.}\ }\textbf {\bibinfo {volume} {98}},\ \bibinfo {pages} {110201} (\bibinfo {year} {2007})}\BibitemShut {NoStop}%
\bibitem [{\citenamefont {Toulouse}\ and\ \citenamefont {Umrigar}(2007)}]{ToulouseUmrigar2007}%
  \BibitemOpen
  \bibfield  {author} {\bibinfo {author} {\bibfnamefont {J.}~\bibnamefont {Toulouse}}\ and\ \bibinfo {author} {\bibfnamefont {C.~J.}\ \bibnamefont {Umrigar}},\ }\bibfield  {title} {\bibinfo {title} {{Optimization of quantum Monte Carlo wave functions by energy minimization}},\ }\href@noop {} {\bibfield  {journal} {\bibinfo  {journal} {J. Chem. Phys.}\ }\textbf {\bibinfo {volume} {126}},\ \bibinfo {pages} {084102} (\bibinfo {year} {2007})}\BibitemShut {NoStop}%
\bibitem [{\citenamefont {Carleo}\ and\ \citenamefont {Troyer}(2017)}]{CarleoTroyer2017}%
  \BibitemOpen
  \bibfield  {author} {\bibinfo {author} {\bibfnamefont {G.}~\bibnamefont {Carleo}}\ and\ \bibinfo {author} {\bibfnamefont {M.}~\bibnamefont {Troyer}},\ }\bibfield  {title} {\bibinfo {title} {{Solving the quantum many-body problem with artificial neural networks}},\ }\href@noop {} {\bibfield  {journal} {\bibinfo  {journal} {Science}\ }\textbf {\bibinfo {volume} {355}},\ \bibinfo {pages} {602} (\bibinfo {year} {2017})}\BibitemShut {NoStop}%
\bibitem [{\citenamefont {Pfau}\ \emph {et~al.}(2020)\citenamefont {Pfau}, \citenamefont {Spencer}, \citenamefont {Matthews},\ and\ \citenamefont {Foulkes}}]{pfau2020}%
  \BibitemOpen
  \bibfield  {author} {\bibinfo {author} {\bibfnamefont {D.}~\bibnamefont {Pfau}}, \bibinfo {author} {\bibfnamefont {J.~S.}\ \bibnamefont {Spencer}}, \bibinfo {author} {\bibfnamefont {A.~G. D.~G.}\ \bibnamefont {Matthews}},\ and\ \bibinfo {author} {\bibfnamefont {W.~M.~C.}\ \bibnamefont {Foulkes}},\ }\bibfield  {title} {\bibinfo {title} {{Ab initio solution of the many-electron Schrödinger equation with deep neural networks}},\ }\href@noop {} {\bibfield  {journal} {\bibinfo  {journal} {Phys. Rev. Res.}\ }\textbf {\bibinfo {volume} {2}},\ \bibinfo {pages} {033429} (\bibinfo {year} {2020})}\BibitemShut {NoStop}%
\bibitem [{\citenamefont {Hermann}\ \emph {et~al.}(2020)\citenamefont {Hermann}, \citenamefont {Sch\"atzle},\ and\ \citenamefont {No\'e}}]{Hermann2020}%
  \BibitemOpen
  \bibfield  {author} {\bibinfo {author} {\bibfnamefont {J.}~\bibnamefont {Hermann}}, \bibinfo {author} {\bibfnamefont {Z.}~\bibnamefont {Sch\"atzle}},\ and\ \bibinfo {author} {\bibfnamefont {F.}~\bibnamefont {No\'e}},\ }\bibfield  {title} {\bibinfo {title} {{Deep-neural-network solution of the electronic Schrödinger equation}},\ }\href@noop {} {\bibfield  {journal} {\bibinfo  {journal} {Nat. Chem.}\ }\textbf {\bibinfo {volume} {12}},\ \bibinfo {pages} {891} (\bibinfo {year} {2020})}\BibitemShut {NoStop}%
\bibitem [{\citenamefont {Temple}(1928)}]{temple1928}%
  \BibitemOpen
  \bibfield  {author} {\bibinfo {author} {\bibfnamefont {G.}~\bibnamefont {Temple}},\ }\bibfield  {title} {\bibinfo {title} {{The Theory of Rayleigh's Principle as Applied to Continuous Systems}},\ }\href@noop {} {\bibfield  {journal} {\bibinfo  {journal} {Proc. R. Soc. Lond. A}\ }\textbf {\bibinfo {volume} {119}},\ \bibinfo {pages} {276} (\bibinfo {year} {1928})}\BibitemShut {NoStop}%
\bibitem [{\citenamefont {Eckart}(1930)}]{Eckart1930}%
  \BibitemOpen
  \bibfield  {author} {\bibinfo {author} {\bibfnamefont {C.}~\bibnamefont {Eckart}},\ }\bibfield  {title} {\bibinfo {title} {{The Theory and Calculation of Screening Constants}},\ }\href@noop {} {\bibfield  {journal} {\bibinfo  {journal} {Phys. Rev.}\ }\textbf {\bibinfo {volume} {36}},\ \bibinfo {pages} {878} (\bibinfo {year} {1930})}\BibitemShut {NoStop}%
\bibitem [{\citenamefont {Weinstein}(1932{\natexlab{a}})}]{Weinstein1932}%
  \BibitemOpen
  \bibfield  {author} {\bibinfo {author} {\bibfnamefont {D.~H.}\ \bibnamefont {Weinstein}},\ }\bibfield  {title} {\bibinfo {title} {{A Lower Limit for the Ground State of the Helium Atom}},\ }\href@noop {} {\bibfield  {journal} {\bibinfo  {journal} {Phys. Rev.}\ }\textbf {\bibinfo {volume} {40}},\ \bibinfo {pages} {797} (\bibinfo {year} {1932}{\natexlab{a}})}\BibitemShut {NoStop}%
\bibitem [{\citenamefont {Weinstein}(1932{\natexlab{b}})}]{Weinstein1932b}%
  \BibitemOpen
  \bibfield  {author} {\bibinfo {author} {\bibfnamefont {D.~H.}\ \bibnamefont {Weinstein}},\ }\bibfield  {title} {\bibinfo {title} {{Lower Limit for the Ground State of the Helium Atom}},\ }\href@noop {} {\bibfield  {journal} {\bibinfo  {journal} {Phys. Rev.}\ }\textbf {\bibinfo {volume} {41}},\ \bibinfo {pages} {839} (\bibinfo {year} {1932}{\natexlab{b}})}\BibitemShut {NoStop}%
\bibitem [{\citenamefont {Weinstein}(1934)}]{Weinstein1934}%
  \BibitemOpen
  \bibfield  {author} {\bibinfo {author} {\bibfnamefont {D.~H.}\ \bibnamefont {Weinstein}},\ }\bibfield  {title} {\bibinfo {title} {{Modified Ritz Method}},\ }\href@noop {} {\bibfield  {journal} {\bibinfo  {journal} {Proc. Nat. Acad. Sci.}\ }\textbf {\bibinfo {volume} {20}},\ \bibinfo {pages} {529} (\bibinfo {year} {1934})}\BibitemShut {NoStop}%
\bibitem [{\citenamefont {MacDonald}(1934)}]{macdonald1934}%
  \BibitemOpen
  \bibfield  {author} {\bibinfo {author} {\bibfnamefont {J.~K.~L.}\ \bibnamefont {MacDonald}},\ }\bibfield  {title} {\bibinfo {title} {{On the Modified Ritz Variation Method}},\ }\href@noop {} {\bibfield  {journal} {\bibinfo  {journal} {Phys. Rev.}\ }\textbf {\bibinfo {volume} {46}},\ \bibinfo {pages} {828} (\bibinfo {year} {1934})}\BibitemShut {NoStop}%
\bibitem [{\citenamefont {Stevenson}(1938)}]{stevenson1938b}%
  \BibitemOpen
  \bibfield  {author} {\bibinfo {author} {\bibfnamefont {A.~F.}\ \bibnamefont {Stevenson}},\ }\bibfield  {title} {\bibinfo {title} {{On the Lower Bounds of Weinstein and Romberg in Quantum Mechanics}},\ }\href@noop {} {\bibfield  {journal} {\bibinfo  {journal} {Phys. Rev.}\ }\textbf {\bibinfo {volume} {53}},\ \bibinfo {pages} {199} (\bibinfo {year} {1938})}\BibitemShut {NoStop}%
\bibitem [{\citenamefont {Stevenson}\ and\ \citenamefont {Crawford}(1938)}]{Stevenson1938}%
  \BibitemOpen
  \bibfield  {author} {\bibinfo {author} {\bibfnamefont {A.~F.}\ \bibnamefont {Stevenson}}\ and\ \bibinfo {author} {\bibfnamefont {M.~F.}\ \bibnamefont {Crawford}},\ }\bibfield  {title} {\bibinfo {title} {{A Lower Limit for the Theoretical Energy of the Normal State of Helium}},\ }\href@noop {} {\bibfield  {journal} {\bibinfo  {journal} {Phys. Rev.}\ }\textbf {\bibinfo {volume} {54}},\ \bibinfo {pages} {375} (\bibinfo {year} {1938})}\BibitemShut {NoStop}%
\bibitem [{\citenamefont {Fr\"oman}\ and\ \citenamefont {Hall}(1961)}]{fromanhall1961}%
  \BibitemOpen
  \bibfield  {author} {\bibinfo {author} {\bibfnamefont {A.}~\bibnamefont {Fr\"oman}}\ and\ \bibinfo {author} {\bibfnamefont {G.~G.}\ \bibnamefont {Hall}},\ }\bibfield  {title} {\bibinfo {title} {{The Accuracy of Atomic Wave Functions and Their Scale}},\ }\href@noop {} {\bibfield  {journal} {\bibinfo  {journal} {J. Mol. Spectrosc.}\ }\textbf {\bibinfo {volume} {7}},\ \bibinfo {pages} {410} (\bibinfo {year} {1961})}\BibitemShut {NoStop}%
\bibitem [{\citenamefont {Conroy}(1967{\natexlab{a}})}]{Conroy1967}%
  \BibitemOpen
  \bibfield  {author} {\bibinfo {author} {\bibfnamefont {H.}~\bibnamefont {Conroy}},\ }\bibfield  {title} {\bibinfo {title} {{Molecular Schr\"odinger Equation. VII. Properties of the Energy Variance Function: The Estimation of Energy Eigenvalues}},\ }\href@noop {} {\bibfield  {journal} {\bibinfo  {journal} {J. Chem. Phys.}\ }\textbf {\bibinfo {volume} {47}},\ \bibinfo {pages} {930} (\bibinfo {year} {1967}{\natexlab{a}})}\BibitemShut {NoStop}%
\bibitem [{\citenamefont {Bartlett}\ \emph {et~al.}(1935)\citenamefont {Bartlett}, \citenamefont {Gibbons},\ and\ \citenamefont {Dunn}}]{Bartlett1935}%
  \BibitemOpen
  \bibfield  {author} {\bibinfo {author} {\bibfnamefont {J.~H.}\ \bibnamefont {Bartlett}, \bibfnamefont {Jr.}}, \bibinfo {author} {\bibfnamefont {J.~J.}\ \bibnamefont {Gibbons}, \bibfnamefont {Jr.}},\ and\ \bibinfo {author} {\bibfnamefont {C.~G.}\ \bibnamefont {Dunn}},\ }\bibfield  {title} {\bibinfo {title} {{The Normal Helium Atom}},\ }\href@noop {} {\bibfield  {journal} {\bibinfo  {journal} {Phys. Rev.}\ }\textbf {\bibinfo {volume} {47}},\ \bibinfo {pages} {679} (\bibinfo {year} {1935})}\BibitemShut {NoStop}%
\bibitem [{\citenamefont {James}\ and\ \citenamefont {Coolidge}(1937)}]{James1937}%
  \BibitemOpen
  \bibfield  {author} {\bibinfo {author} {\bibfnamefont {H.~M.}\ \bibnamefont {James}}\ and\ \bibinfo {author} {\bibfnamefont {A.~S.}\ \bibnamefont {Coolidge}},\ }\bibfield  {title} {\bibinfo {title} {{Criteria of Goodness for Approximate Wave Functions}},\ }\href@noop {} {\bibfield  {journal} {\bibinfo  {journal} {Phys. Rev.}\ }\textbf {\bibinfo {volume} {51}},\ \bibinfo {pages} {860} (\bibinfo {year} {1937})}\BibitemShut {NoStop}%
\bibitem [{\citenamefont {Fu}\ \emph {et~al.}(2024)\citenamefont {Fu}, \citenamefont {Ren},\ and\ \citenamefont {Chen}}]{Fu2024}%
  \BibitemOpen
  \bibfield  {author} {\bibinfo {author} {\bibfnamefont {W.}~\bibnamefont {Fu}}, \bibinfo {author} {\bibfnamefont {W.}~\bibnamefont {Ren}},\ and\ \bibinfo {author} {\bibfnamefont {J.}~\bibnamefont {Chen}},\ }\bibfield  {title} {\bibinfo {title} {{Variance extrapolation method for neural-network variational Monte Carlo}},\ }\href@noop {} {\bibfield  {journal} {\bibinfo  {journal} {Mach. Learn.: Sci. Technol.}\ }\textbf {\bibinfo {volume} {5}},\ \bibinfo {pages} {015016} (\bibinfo {year} {2024})}\BibitemShut {NoStop}%
\bibitem [{\citenamefont {Chen}\ and\ \citenamefont {Heyl}(2024)}]{Chen2024}%
  \BibitemOpen
  \bibfield  {author} {\bibinfo {author} {\bibfnamefont {A.}~\bibnamefont {Chen}}\ and\ \bibinfo {author} {\bibfnamefont {M.}~\bibnamefont {Heyl}},\ }\bibfield  {title} {\bibinfo {title} {{Empowering deep neural quantum states through efficient optimization}},\ }\href@noop {} {\bibfield  {journal} {\bibinfo  {journal} {Nat. Phys.}\ }\textbf {\bibinfo {volume} {20}},\ \bibinfo {pages} {1476} (\bibinfo {year} {2024})}\BibitemShut {NoStop}%
\bibitem [{\citenamefont {James}\ and\ \citenamefont {Yost}(1938)}]{JamesYost1938}%
  \BibitemOpen
  \bibfield  {author} {\bibinfo {author} {\bibfnamefont {H.~M.}\ \bibnamefont {James}}\ and\ \bibinfo {author} {\bibfnamefont {F.~L.}\ \bibnamefont {Yost}},\ }\bibfield  {title} {\bibinfo {title} {{Wave Functions for $1s2s\,^3S$ ${\textrm{Li}^+}$}},\ }\href@noop {} {\bibfield  {journal} {\bibinfo  {journal} {Phys. Rev.}\ }\textbf {\bibinfo {volume} {54}},\ \bibinfo {pages} {646} (\bibinfo {year} {1938})}\BibitemShut {NoStop}%
\bibitem [{\citenamefont {James}\ and\ \citenamefont {Coolidge}(1939)}]{James1939}%
  \BibitemOpen
  \bibfield  {author} {\bibinfo {author} {\bibfnamefont {H.~M.}\ \bibnamefont {James}}\ and\ \bibinfo {author} {\bibfnamefont {A.~S.}\ \bibnamefont {Coolidge}},\ }\bibfield  {title} {\bibinfo {title} {{Symmetry Properties of Variational Functions}},\ }\href@noop {} {\bibfield  {journal} {\bibinfo  {journal} {Phys. Rev.}\ }\textbf {\bibinfo {volume} {55}},\ \bibinfo {pages} {873} (\bibinfo {year} {1939})}\BibitemShut {NoStop}%
\bibitem [{\citenamefont {Williamson}(1942)}]{Williamson1942}%
  \BibitemOpen
  \bibfield  {author} {\bibinfo {author} {\bibfnamefont {R.~E.}\ \bibnamefont {Williamson}},\ }\bibfield  {title} {\bibinfo {title} {{The Wave Function for the Negative Hydrogen Ion}},\ }\href@noop {} {\bibfield  {journal} {\bibinfo  {journal} {Phys. Rev.}\ }\textbf {\bibinfo {volume} {62}},\ \bibinfo {pages} {538} (\bibinfo {year} {1942})}\BibitemShut {NoStop}%
\bibitem [{\citenamefont {Frost}(1942)}]{Frost1942}%
  \BibitemOpen
  \bibfield  {author} {\bibinfo {author} {\bibfnamefont {A.~A.}\ \bibnamefont {Frost}},\ }\bibfield  {title} {\bibinfo {title} {{The Approximate Solution of Schr\"odinger Equations by a Least Squares Method}},\ }\href@noop {} {\bibfield  {journal} {\bibinfo  {journal} {J. Chem. Phys.}\ }\textbf {\bibinfo {volume} {10}},\ \bibinfo {pages} {240} (\bibinfo {year} {1942})}\BibitemShut {NoStop}%
\bibitem [{\citenamefont {von Mohrenstein}(1953)}]{Mohrenstein1953}%
  \BibitemOpen
  \bibfield  {author} {\bibinfo {author} {\bibfnamefont {A.}~\bibnamefont {von Mohrenstein}},\ }\bibfield  {title} {\bibinfo {title} {{Numerische Aufl\"osung der Schr\"odinger-Gleichung}},\ }\href@noop {} {\bibfield  {journal} {\bibinfo  {journal} {Z. Phys.}\ }\textbf {\bibinfo {volume} {134}},\ \bibinfo {pages} {488} (\bibinfo {year} {1953})}\BibitemShut {NoStop}%
\bibitem [{\citenamefont {Bartlett}(1955)}]{Bartlett1955}%
  \BibitemOpen
  \bibfield  {author} {\bibinfo {author} {\bibfnamefont {J.~H.}\ \bibnamefont {Bartlett}},\ }\bibfield  {title} {\bibinfo {title} {{Helium Wave Equation}},\ }\href@noop {} {\bibfield  {journal} {\bibinfo  {journal} {Physical Review}\ }\textbf {\bibinfo {volume} {98}},\ \bibinfo {pages} {1067} (\bibinfo {year} {1955})}\BibitemShut {NoStop}%
\bibitem [{\citenamefont {Preuss}\ and\ \citenamefont {Trefftz}(1957)}]{Preuss1957}%
  \BibitemOpen
  \bibfield  {author} {\bibinfo {author} {\bibfnamefont {H.}~\bibnamefont {Preuss}}\ and\ \bibinfo {author} {\bibfnamefont {E.}~\bibnamefont {Trefftz}},\ }\bibfield  {title} {\bibinfo {title} {{$\delta^2$ Test for Atomic Wave Functions, Applied on {\textrm{H}}}},\ }\href@noop {} {\bibfield  {journal} {\bibinfo  {journal} {Phys. Rev.}\ }\textbf {\bibinfo {volume} {107}},\ \bibinfo {pages} {1282} (\bibinfo {year} {1957})}\BibitemShut {NoStop}%
\bibitem [{\citenamefont {Preuss}(1958)}]{Preuss1958}%
  \BibitemOpen
  \bibfield  {author} {\bibinfo {author} {\bibfnamefont {H.}~\bibnamefont {Preuss}},\ }\bibfield  {title} {\bibinfo {title} {{\"{U}ber die G\"ute von N\"aherungsl\"osungen}},\ }\href@noop {} {\bibfield  {journal} {\bibinfo  {journal} {Z. Naturforsch.}\ }\textbf {\bibinfo {volume} {13a}},\ \bibinfo {pages} {439} (\bibinfo {year} {1958})}\BibitemShut {NoStop}%
\bibitem [{\citenamefont {Preuss}(1961)}]{Preuss1961}%
  \BibitemOpen
  \bibfield  {author} {\bibinfo {author} {\bibfnamefont {H.}~\bibnamefont {Preuss}},\ }\bibfield  {title} {\bibinfo {title} {{Die Bedeutung des Energievergleiches f\"ur die G\"ute einer N\"aherungsl\"osung}},\ }\href@noop {} {\bibfield  {journal} {\bibinfo  {journal} {Z. Naturforsch.}\ }\textbf {\bibinfo {volume} {16a}},\ \bibinfo {pages} {598} (\bibinfo {year} {1961})}\BibitemShut {NoStop}%
\bibitem [{\citenamefont {Frost}\ \emph {et~al.}(1960)\citenamefont {Frost}, \citenamefont {Kellogg},\ and\ \citenamefont {Curtis}}]{Frost1960}%
  \BibitemOpen
  \bibfield  {author} {\bibinfo {author} {\bibfnamefont {A.~A.}\ \bibnamefont {Frost}}, \bibinfo {author} {\bibfnamefont {R.~E.}\ \bibnamefont {Kellogg}},\ and\ \bibinfo {author} {\bibfnamefont {E.~C.}\ \bibnamefont {Curtis}},\ }\bibfield  {title} {\bibinfo {title} {{Local-Energy Method in Electronic Energy Calculations}},\ }\href@noop {} {\bibfield  {journal} {\bibinfo  {journal} {Rev. Mod. Phys.}\ }\textbf {\bibinfo {volume} {32}},\ \bibinfo {pages} {313} (\bibinfo {year} {1960})}\BibitemShut {NoStop}%
\bibitem [{\citenamefont {Frost}\ \emph {et~al.}(1961)\citenamefont {Frost}, \citenamefont {Kellogg}, \citenamefont {Gimarc},\ and\ \citenamefont {Scargle}}]{Frost1961}%
  \BibitemOpen
  \bibfield  {author} {\bibinfo {author} {\bibfnamefont {A.~A.}\ \bibnamefont {Frost}}, \bibinfo {author} {\bibfnamefont {R.~E.}\ \bibnamefont {Kellogg}}, \bibinfo {author} {\bibfnamefont {B.~M.}\ \bibnamefont {Gimarc}},\ and\ \bibinfo {author} {\bibfnamefont {J.~D.}\ \bibnamefont {Scargle}},\ }\bibfield  {title} {\bibinfo {title} {{Least-Squares Local-Energy Method for Molecular Energy Calculations Using Gauss Quadrature Points}},\ }\href@noop {} {\bibfield  {journal} {\bibinfo  {journal} {J. Chem. Phys.}\ }\textbf {\bibinfo {volume} {35}},\ \bibinfo {pages} {827} (\bibinfo {year} {1961})}\BibitemShut {NoStop}%
\bibitem [{\citenamefont {Gimarc}\ and\ \citenamefont {Frost}(1963{\natexlab{a}})}]{Gimarc1963a}%
  \BibitemOpen
  \bibfield  {author} {\bibinfo {author} {\bibfnamefont {B.~M.}\ \bibnamefont {Gimarc}}\ and\ \bibinfo {author} {\bibfnamefont {A.~A.}\ \bibnamefont {Frost}},\ }\bibfield  {title} {\bibinfo {title} {{Energy of the Lithium Atom by the Least-Squares Local Energy Method}},\ }\href@noop {} {\bibfield  {journal} {\bibinfo  {journal} {J. Chem. Phys.}\ }\textbf {\bibinfo {volume} {39}},\ \bibinfo {pages} {1698} (\bibinfo {year} {1963}{\natexlab{a}})}\BibitemShut {NoStop}%
\bibitem [{\citenamefont {Gimarc}\ and\ \citenamefont {Frost}(1963{\natexlab{b}})}]{Gimarc1963b}%
  \BibitemOpen
  \bibfield  {author} {\bibinfo {author} {\bibfnamefont {B.~M.}\ \bibnamefont {Gimarc}}\ and\ \bibinfo {author} {\bibfnamefont {A.~A.}\ \bibnamefont {Frost}},\ }\bibfield  {title} {\bibinfo {title} {{Energies of the Lowest Singlet $S$ and Triplet $S$ States of Helium by the Local Energy Method}},\ }\href@noop {} {\bibfield  {journal} {\bibinfo  {journal} {Theoret. Chim. Acta}\ }\textbf {\bibinfo {volume} {1}},\ \bibinfo {pages} {87} (\bibinfo {year} {1963}{\natexlab{b}})}\BibitemShut {NoStop}%
\bibitem [{\citenamefont {Harriss}\ and\ \citenamefont {Frost}(1964)}]{Harriss1964}%
  \BibitemOpen
  \bibfield  {author} {\bibinfo {author} {\bibfnamefont {D.~K.}\ \bibnamefont {Harriss}}\ and\ \bibinfo {author} {\bibfnamefont {A.~A.}\ \bibnamefont {Frost}},\ }\bibfield  {title} {\bibinfo {title} {{Electronic Energy of the Hydrogen Molecule Ground State by the Local-Energy Method}},\ }\href@noop {} {\bibfield  {journal} {\bibinfo  {journal} {J. Chem. Phys.}\ }\textbf {\bibinfo {volume} {40}},\ \bibinfo {pages} {204} (\bibinfo {year} {1964})}\BibitemShut {NoStop}%
\bibitem [{\citenamefont {Goodisman}\ and\ \citenamefont {Secrest}(1964)}]{GoodismanSecrest1964}%
  \BibitemOpen
  \bibfield  {author} {\bibinfo {author} {\bibfnamefont {J.}~\bibnamefont {Goodisman}}\ and\ \bibinfo {author} {\bibfnamefont {D.}~\bibnamefont {Secrest}},\ }\bibfield  {title} {\bibinfo {title} {{Weinstein Calculation on Hydrogen Molecular Ion}},\ }\href@noop {} {\bibfield  {journal} {\bibinfo  {journal} {J. Chem. Phys.}\ }\textbf {\bibinfo {volume} {41}},\ \bibinfo {pages} {3610} (\bibinfo {year} {1964})}\BibitemShut {NoStop}%
\bibitem [{\citenamefont {Goodisman}(1964)}]{Goodisman1964}%
  \BibitemOpen
  \bibfield  {author} {\bibinfo {author} {\bibfnamefont {J.}~\bibnamefont {Goodisman}},\ }\bibfield  {title} {\bibinfo {title} {{Numerical Calculation of ${\textrm{H}_2^+}$ and ${\textrm{H}_2}$ Wavefunctions}},\ }\href@noop {} {\bibfield  {journal} {\bibinfo  {journal} {J. Chem. Phys.}\ }\textbf {\bibinfo {volume} {41}},\ \bibinfo {pages} {3889} (\bibinfo {year} {1964})}\BibitemShut {NoStop}%
\bibitem [{\citenamefont {Conroy}(1964{\natexlab{a}})}]{Conroy1964comm}%
  \BibitemOpen
  \bibfield  {author} {\bibinfo {author} {\bibfnamefont {H.}~\bibnamefont {Conroy}},\ }\bibfield  {title} {\bibinfo {title} {{Potential Energy Surfaces for the ${\textrm{H}_3^+}$ Molecule-Ion}},\ }\href@noop {} {\bibfield  {journal} {\bibinfo  {journal} {J. Chem. Phys.}\ }\textbf {\bibinfo {volume} {40}},\ \bibinfo {pages} {603} (\bibinfo {year} {1964}{\natexlab{a}})}\BibitemShut {NoStop}%
\bibitem [{\citenamefont {Conroy}(1964{\natexlab{b}})}]{Conroy1964I}%
  \BibitemOpen
  \bibfield  {author} {\bibinfo {author} {\bibfnamefont {H.}~\bibnamefont {Conroy}},\ }\bibfield  {title} {\bibinfo {title} {{Molecular Schrödinger Equation. I. One-Electron Solutions}},\ }\href@noop {} {\bibfield  {journal} {\bibinfo  {journal} {J. Chem. Phys.}\ }\textbf {\bibinfo {volume} {41}},\ \bibinfo {pages} {1327} (\bibinfo {year} {1964}{\natexlab{b}})}\BibitemShut {NoStop}%
\bibitem [{\citenamefont {Conroy}(1964{\natexlab{c}})}]{Conroy1964a}%
  \BibitemOpen
  \bibfield  {author} {\bibinfo {author} {\bibfnamefont {H.}~\bibnamefont {Conroy}},\ }\bibfield  {title} {\bibinfo {title} {{Molecular Schrödinger Equation. II. Monte Carlo Evaluation of Integrals}},\ }\href@noop {} {\bibfield  {journal} {\bibinfo  {journal} {J. Chem. Phys.}\ }\textbf {\bibinfo {volume} {41}},\ \bibinfo {pages} {1331} (\bibinfo {year} {1964}{\natexlab{c}})}\BibitemShut {NoStop}%
\bibitem [{\citenamefont {Conroy}(1964{\natexlab{d}})}]{Conroy1964b}%
  \BibitemOpen
  \bibfield  {author} {\bibinfo {author} {\bibfnamefont {H.}~\bibnamefont {Conroy}},\ }\bibfield  {title} {\bibinfo {title} {{Molecular Schrödinger Equation. III. Calculation of Ground-State Energies by Extrapolation}},\ }\href@noop {} {\bibfield  {journal} {\bibinfo  {journal} {J. Chem. Phys.}\ }\textbf {\bibinfo {volume} {41}},\ \bibinfo {pages} {1336} (\bibinfo {year} {1964}{\natexlab{d}})}\BibitemShut {NoStop}%
\bibitem [{\citenamefont {Conroy}(1964{\natexlab{e}})}]{Conroy1964IV}%
  \BibitemOpen
  \bibfield  {author} {\bibinfo {author} {\bibfnamefont {H.}~\bibnamefont {Conroy}},\ }\bibfield  {title} {\bibinfo {title} {{Molecular Schr\"odinger Equation. IV. Results for One- and Two-Electron Systems}},\ }\href@noop {} {\bibfield  {journal} {\bibinfo  {journal} {J. Chem. Phys.}\ }\textbf {\bibinfo {volume} {41}},\ \bibinfo {pages} {1341} (\bibinfo {year} {1964}{\natexlab{e}})}\BibitemShut {NoStop}%
\bibitem [{\citenamefont {Conroy}\ and\ \citenamefont {Bruner}(1965)}]{ConroyBruner1965}%
  \BibitemOpen
  \bibfield  {author} {\bibinfo {author} {\bibfnamefont {H.}~\bibnamefont {Conroy}}\ and\ \bibinfo {author} {\bibfnamefont {B.~L.}\ \bibnamefont {Bruner}},\ }\bibfield  {title} {\bibinfo {title} {{Energy Surface for the Linear ${\textrm{H}_3}$ System}},\ }\href@noop {} {\bibfield  {journal} {\bibinfo  {journal} {J. Chem. Phys.}\ }\textbf {\bibinfo {volume} {42}},\ \bibinfo {pages} {4047} (\bibinfo {year} {1965})}\BibitemShut {NoStop}%
\bibitem [{\citenamefont {Conroy}(1967{\natexlab{b}})}]{ConroyV1967}%
  \BibitemOpen
  \bibfield  {author} {\bibinfo {author} {\bibfnamefont {H.}~\bibnamefont {Conroy}},\ }\bibfield  {title} {\bibinfo {title} {{Molecular Schrödinger Equation. V. A Correlated Polyelectronic Wavefunction}},\ }\href@noop {} {\bibfield  {journal} {\bibinfo  {journal} {J. Chem. Phys.}\ }\textbf {\bibinfo {volume} {47}},\ \bibinfo {pages} {912} (\bibinfo {year} {1967}{\natexlab{b}})}\BibitemShut {NoStop}%
\bibitem [{\citenamefont {Conroy}\ and\ \citenamefont {Bruner}(1967)}]{ConroyBruner1967}%
  \BibitemOpen
  \bibfield  {author} {\bibinfo {author} {\bibfnamefont {H.}~\bibnamefont {Conroy}}\ and\ \bibinfo {author} {\bibfnamefont {B.~L.}\ \bibnamefont {Bruner}},\ }\bibfield  {title} {\bibinfo {title} {{Molecular Schrödinger Equation. VI. Results for ${\textrm{H}_3}$ and Other Simple Systems}},\ }\href@noop {} {\bibfield  {journal} {\bibinfo  {journal} {J. Chem. Phys.}\ }\textbf {\bibinfo {volume} {47}},\ \bibinfo {pages} {921} (\bibinfo {year} {1967})}\BibitemShut {NoStop}%
\bibitem [{\citenamefont {Conroy}(1967{\natexlab{c}})}]{Conroy1967VIII}%
  \BibitemOpen
  \bibfield  {author} {\bibinfo {author} {\bibfnamefont {H.}~\bibnamefont {Conroy}},\ }\bibfield  {title} {\bibinfo {title} {{Molecular Schrödinger Equation. VIII. A New Method for the Evaluation of Multidimensional Integrals}},\ }\href@noop {} {\bibfield  {journal} {\bibinfo  {journal} {J. Chem. Phys.}\ }\textbf {\bibinfo {volume} {47}},\ \bibinfo {pages} {5307} (\bibinfo {year} {1967}{\natexlab{c}})}\BibitemShut {NoStop}%
\bibitem [{\citenamefont {Conroy}\ and\ \citenamefont {Malli}(1969)}]{ConroyMalli1969}%
  \BibitemOpen
  \bibfield  {author} {\bibinfo {author} {\bibfnamefont {H.}~\bibnamefont {Conroy}}\ and\ \bibinfo {author} {\bibfnamefont {G.}~\bibnamefont {Malli}},\ }\bibfield  {title} {\bibinfo {title} {{Molecular Schrödinger Equation. IX. Square and Rectangular States of ${\textrm{H}_4}$ and the Molecular Ions ${\textrm{H}_4^{3+}}$ and ${\textrm{H}_4^{2+}}$}},\ }\href@noop {} {\bibfield  {journal} {\bibinfo  {journal} {J. Chem. Phys.}\ }\textbf {\bibinfo {volume} {50}},\ \bibinfo {pages} {5049} (\bibinfo {year} {1969})}\BibitemShut {NoStop}%
\bibitem [{\citenamefont {Conroy}(1969)}]{Conroy1969X}%
  \BibitemOpen
  \bibfield  {author} {\bibinfo {author} {\bibfnamefont {H.}~\bibnamefont {Conroy}},\ }\bibfield  {title} {\bibinfo {title} {{Molecular Schrödinger Equation. X. Potential Surfaces for Ground and Excited States of Isosceles ${\textrm{H}_3^{++}}$ and ${\textrm{H}_3^{+}}$}},\ }\href@noop {} {\bibfield  {journal} {\bibinfo  {journal} {J. Chem. Phys.}\ }\textbf {\bibinfo {volume} {51}},\ \bibinfo {pages} {3979} (\bibinfo {year} {1969})}\BibitemShut {NoStop}%
\bibitem [{\citenamefont {Fraga}\ and\ \citenamefont {Birss}(1964)}]{FragaBirss1964}%
  \BibitemOpen
  \bibfield  {author} {\bibinfo {author} {\bibfnamefont {S.}~\bibnamefont {Fraga}}\ and\ \bibinfo {author} {\bibfnamefont {F.~W.}\ \bibnamefont {Birss}},\ }\bibfield  {title} {\bibinfo {title} {{Self-Consistent-Field Theory. III. General Treatment for Excited States}},\ }\href@noop {} {\bibfield  {journal} {\bibinfo  {journal} {J. Chem. Phys.}\ }\textbf {\bibinfo {volume} {40}},\ \bibinfo {pages} {3207} (\bibinfo {year} {1964})}\BibitemShut {NoStop}%
\bibitem [{\citenamefont {Stanton}\ and\ \citenamefont {Taylor}(1966)}]{Stanton1966}%
  \BibitemOpen
  \bibfield  {author} {\bibinfo {author} {\bibfnamefont {R.~E.}\ \bibnamefont {Stanton}}\ and\ \bibinfo {author} {\bibfnamefont {R.~L.}\ \bibnamefont {Taylor}},\ }\bibfield  {title} {\bibinfo {title} {{Mathematical Properties of Frost's Local-Energy Method}},\ }\href@noop {} {\bibfield  {journal} {\bibinfo  {journal} {J. Chem. Phys.}\ }\textbf {\bibinfo {volume} {45}},\ \bibinfo {pages} {565} (\bibinfo {year} {1966})}\BibitemShut {NoStop}%
\bibitem [{\citenamefont {Harriss}\ and\ \citenamefont {Roubal}(1968)}]{Harriss1968}%
  \BibitemOpen
  \bibfield  {author} {\bibinfo {author} {\bibfnamefont {D.~K.}\ \bibnamefont {Harriss}}\ and\ \bibinfo {author} {\bibfnamefont {R.~K.}\ \bibnamefont {Roubal}},\ }\bibfield  {title} {\bibinfo {title} {{Nonlinear Parameters in the Least-Squares Local Energy Method}},\ }\href@noop {} {\bibfield  {journal} {\bibinfo  {journal} {Theoret. Chim. Acta}\ }\textbf {\bibinfo {volume} {9}},\ \bibinfo {pages} {303} (\bibinfo {year} {1968})}\BibitemShut {NoStop}%
\bibitem [{\citenamefont {Harriss}\ and\ \citenamefont {Solev}(1975)}]{Harriss1975}%
  \BibitemOpen
  \bibfield  {author} {\bibinfo {author} {\bibfnamefont {D.~K.}\ \bibnamefont {Harriss}}\ and\ \bibinfo {author} {\bibfnamefont {I.~G.}\ \bibnamefont {Solev}},\ }\bibfield  {title} {\bibinfo {title} {{On the Solution of the Least-Squares Local Energy Variance Minimization Equations}},\ }\href@noop {} {\bibfield  {journal} {\bibinfo  {journal} {Int. J. Quantum Chem.}\ }\textbf {\bibinfo {volume} {9}},\ \bibinfo {pages} {975} (\bibinfo {year} {1975})}\BibitemShut {NoStop}%
\bibitem [{\citenamefont {Goodisman}(1965)}]{Goodisman1965}%
  \BibitemOpen
  \bibfield  {author} {\bibinfo {author} {\bibfnamefont {J.}~\bibnamefont {Goodisman}},\ }\bibfield  {title} {\bibinfo {title} {{Calculation of Energies of Excited States}},\ }\href@noop {} {\bibfield  {journal} {\bibinfo  {journal} {J. Chem. Phys.}\ }\textbf {\bibinfo {volume} {43}},\ \bibinfo {pages} {2806} (\bibinfo {year} {1965})}\BibitemShut {NoStop}%
\bibitem [{\citenamefont {Goodisman}\ and\ \citenamefont {Secrest}(1966)}]{GoodismanSecrest1966}%
  \BibitemOpen
  \bibfield  {author} {\bibinfo {author} {\bibfnamefont {J.}~\bibnamefont {Goodisman}}\ and\ \bibinfo {author} {\bibfnamefont {D.}~\bibnamefont {Secrest}},\ }\bibfield  {title} {\bibinfo {title} {{Use of Numerical Integration in the Computation of the Expectation Value of $H^2$ with Applications to ${\textrm{H}_2}$}},\ }\href@noop {} {\bibfield  {journal} {\bibinfo  {journal} {J. Chem. Phys.}\ }\textbf {\bibinfo {volume} {45}},\ \bibinfo {pages} {1515} (\bibinfo {year} {1966})}\BibitemShut {NoStop}%
\bibitem [{\citenamefont {Goodisman}(1966)}]{Goodisman1966}%
  \BibitemOpen
  \bibfield  {author} {\bibinfo {author} {\bibfnamefont {J.}~\bibnamefont {Goodisman}},\ }\bibfield  {title} {\bibinfo {title} {{Minimization of the Width as an Alternative to the Conventional Variation Method}},\ }\href@noop {} {\bibfield  {journal} {\bibinfo  {journal} {J. Chem. Phys.}\ }\textbf {\bibinfo {volume} {45}},\ \bibinfo {pages} {3659} (\bibinfo {year} {1966})}\BibitemShut {NoStop}%
\bibitem [{\citenamefont {Goodisman}(1967)}]{Goodisman1967}%
  \BibitemOpen
  \bibfield  {author} {\bibinfo {author} {\bibfnamefont {J.}~\bibnamefont {Goodisman}},\ }\bibfield  {title} {\bibinfo {title} {{Weinstein Calculations for Excited States}},\ }\href@noop {} {\bibfield  {journal} {\bibinfo  {journal} {J. Chem. Phys.}\ }\textbf {\bibinfo {volume} {47}},\ \bibinfo {pages} {5247} (\bibinfo {year} {1967})}\BibitemShut {NoStop}%
\bibitem [{\citenamefont {Lloyd}\ and\ \citenamefont {Delves}(1969)}]{Lloyd1969}%
  \BibitemOpen
  \bibfield  {author} {\bibinfo {author} {\bibfnamefont {M.~H.}\ \bibnamefont {Lloyd}}\ and\ \bibinfo {author} {\bibfnamefont {L.~M.}\ \bibnamefont {Delves}},\ }\bibfield  {title} {\bibinfo {title} {{On the Least Squares Procedure for Atomic Calculations}},\ }\href@noop {} {\bibfield  {journal} {\bibinfo  {journal} {Int. J. Quantum Chem.}\ }\textbf {\bibinfo {volume} {3}},\ \bibinfo {pages} {169} (\bibinfo {year} {1969})}\BibitemShut {NoStop}%
\bibitem [{\citenamefont {Messmer}(1969)}]{Messmer1969}%
  \BibitemOpen
  \bibfield  {author} {\bibinfo {author} {\bibfnamefont {R.~P.}\ \bibnamefont {Messmer}},\ }\bibfield  {title} {\bibinfo {title} {{On a Variational Method for Determining Excited State Wave Functions}},\ }\href@noop {} {\bibfield  {journal} {\bibinfo  {journal} {Theoret. Chim. Acta}\ }\textbf {\bibinfo {volume} {14}},\ \bibinfo {pages} {319} (\bibinfo {year} {1969})}\BibitemShut {NoStop}%
\bibitem [{\citenamefont {Choi}\ \emph {et~al.}(1970)\citenamefont {Choi}, \citenamefont {Lebeda},\ and\ \citenamefont {Messmer}}]{Choi1970}%
  \BibitemOpen
  \bibfield  {author} {\bibinfo {author} {\bibfnamefont {J.~H.}\ \bibnamefont {Choi}}, \bibinfo {author} {\bibfnamefont {C.~F.}\ \bibnamefont {Lebeda}},\ and\ \bibinfo {author} {\bibfnamefont {R.~P.}\ \bibnamefont {Messmer}},\ }\bibfield  {title} {\bibinfo {title} {{Variational principle for excited states: Exact formulation and other extensions}},\ }\href@noop {} {\bibfield  {journal} {\bibinfo  {journal} {Chem. Phys. Lett.}\ }\textbf {\bibinfo {volume} {5}},\ \bibinfo {pages} {503} (\bibinfo {year} {1970})}\BibitemShut {NoStop}%
\bibitem [{\citenamefont {Cohen}\ and\ \citenamefont {Feldmann}(1970{\natexlab{a}})}]{cohenfeldmann1970b}%
  \BibitemOpen
  \bibfield  {author} {\bibinfo {author} {\bibfnamefont {M.}~\bibnamefont {Cohen}}\ and\ \bibinfo {author} {\bibfnamefont {T.}~\bibnamefont {Feldmann}},\ }\bibfield  {title} {\bibinfo {title} {{On a Variational Method for Calculating Excited-State Wavefunctions}},\ }\href@noop {} {\bibfield  {journal} {\bibinfo  {journal} {Chem. Phys. Lett.}\ }\textbf {\bibinfo {volume} {6}},\ \bibinfo {pages} {43} (\bibinfo {year} {1970}{\natexlab{a}})}\BibitemShut {NoStop}%
\bibitem [{\citenamefont {Weinhold}(1970)}]{weinhold1970}%
  \BibitemOpen
  \bibfield  {author} {\bibinfo {author} {\bibfnamefont {F.}~\bibnamefont {Weinhold}},\ }\bibfield  {title} {\bibinfo {title} {{Criteria of Accuracy of Approximate Wavefunctions}},\ }\href@noop {} {\bibfield  {journal} {\bibinfo  {journal} {J. Math. Phys.}\ }\textbf {\bibinfo {volume} {11}},\ \bibinfo {pages} {2127} (\bibinfo {year} {1970})}\BibitemShut {NoStop}%
\bibitem [{\citenamefont {Cohen}\ and\ \citenamefont {Feldmann}(1969)}]{CohenFeldmann1969}%
  \BibitemOpen
  \bibfield  {author} {\bibinfo {author} {\bibfnamefont {M.}~\bibnamefont {Cohen}}\ and\ \bibinfo {author} {\bibfnamefont {T.}~\bibnamefont {Feldmann}},\ }\bibfield  {title} {\bibinfo {title} {{Lower Bounds to Eigenvalues}},\ }\href@noop {} {\bibfield  {journal} {\bibinfo  {journal} {Can. J. Phys.}\ }\textbf {\bibinfo {volume} {47}},\ \bibinfo {pages} {1877} (\bibinfo {year} {1969})}\BibitemShut {NoStop}%
\bibitem [{\citenamefont {Cohen}\ and\ \citenamefont {Feldmann}(1970{\natexlab{b}})}]{cohenfeldmann1970}%
  \BibitemOpen
  \bibfield  {author} {\bibinfo {author} {\bibfnamefont {M.}~\bibnamefont {Cohen}}\ and\ \bibinfo {author} {\bibfnamefont {T.}~\bibnamefont {Feldmann}},\ }\bibfield  {title} {\bibinfo {title} {{Upper bounds to the overlap between approximate and exact wave functions}},\ }\href@noop {} {\bibfield  {journal} {\bibinfo  {journal} {Can. J. Phys.}\ }\textbf {\bibinfo {volume} {48}},\ \bibinfo {pages} {1681} (\bibinfo {year} {1970}{\natexlab{b}})}\BibitemShut {NoStop}%
\bibitem [{\citenamefont {McMillan}(1965)}]{McMillan1965}%
  \BibitemOpen
  \bibfield  {author} {\bibinfo {author} {\bibfnamefont {W.~L.}\ \bibnamefont {McMillan}},\ }\bibfield  {title} {\bibinfo {title} {{Ground State of Liquid ${\textrm{He}^4}$}},\ }\href@noop {} {\bibfield  {journal} {\bibinfo  {journal} {Phys. Rev.}\ }\textbf {\bibinfo {volume} {138}},\ \bibinfo {pages} {A442} (\bibinfo {year} {1965})}\BibitemShut {NoStop}%
\bibitem [{\citenamefont {Coldwell}(1977)}]{Coldwell1977}%
  \BibitemOpen
  \bibfield  {author} {\bibinfo {author} {\bibfnamefont {R.~L.}\ \bibnamefont {Coldwell}},\ }\bibfield  {title} {\bibinfo {title} {Zero monte carlo error or quantum mechanics is easier},\ }\href@noop {} {\bibfield  {journal} {\bibinfo  {journal} {Int. J. Quantum Chem.}\ }\textbf {\bibinfo {volume} {12}},\ \bibinfo {pages} {215} (\bibinfo {year} {1977})}\BibitemShut {NoStop}%
\bibitem [{\citenamefont {Foulkes}\ \emph {et~al.}(2001)\citenamefont {Foulkes}, \citenamefont {Mitas}, \citenamefont {Needs},\ and\ \citenamefont {Rajagopal}}]{Foulkes2001}%
  \BibitemOpen
  \bibfield  {author} {\bibinfo {author} {\bibfnamefont {W.~M.~C.}\ \bibnamefont {Foulkes}}, \bibinfo {author} {\bibfnamefont {L.}~\bibnamefont {Mitas}}, \bibinfo {author} {\bibfnamefont {R.~J.}\ \bibnamefont {Needs}},\ and\ \bibinfo {author} {\bibfnamefont {G.}~\bibnamefont {Rajagopal}},\ }\bibfield  {title} {\bibinfo {title} {{Quantum Monte Carlo simulations of solids}},\ }\href@noop {} {\bibfield  {journal} {\bibinfo  {journal} {Rev. Mod. Phys.}\ }\textbf {\bibinfo {volume} {73}},\ \bibinfo {pages} {33} (\bibinfo {year} {2001})}\BibitemShut {NoStop}%
\bibitem [{\citenamefont {Austin}\ \emph {et~al.}(2012)\citenamefont {Austin}, \citenamefont {Zubarev},\ and\ \citenamefont {{W. A. Lester Jr.}}}]{Austin2012}%
  \BibitemOpen
  \bibfield  {author} {\bibinfo {author} {\bibfnamefont {B.~M.}\ \bibnamefont {Austin}}, \bibinfo {author} {\bibfnamefont {D.~Y.}\ \bibnamefont {Zubarev}},\ and\ \bibinfo {author} {\bibnamefont {{W. A. Lester Jr.}}},\ }\bibfield  {title} {\bibinfo {title} {{Quantum Monte Carlo and related approaches}},\ }\href@noop {} {\bibfield  {journal} {\bibinfo  {journal} {Chem. Rev.}\ }\textbf {\bibinfo {volume} {112}},\ \bibinfo {pages} {263} (\bibinfo {year} {2012})}\BibitemShut {NoStop}%
\bibitem [{\citenamefont {Alexander}\ \emph {et~al.}(1991)\citenamefont {Alexander}, \citenamefont {Coldwell}, \citenamefont {Monkhorst},\ and\ \citenamefont {{Morgan III}}}]{Alexander1991}%
  \BibitemOpen
  \bibfield  {author} {\bibinfo {author} {\bibfnamefont {S.~A.}\ \bibnamefont {Alexander}}, \bibinfo {author} {\bibfnamefont {R.~L.}\ \bibnamefont {Coldwell}}, \bibinfo {author} {\bibfnamefont {H.~J.}\ \bibnamefont {Monkhorst}},\ and\ \bibinfo {author} {\bibfnamefont {J.~D.}\ \bibnamefont {{Morgan III}}},\ }\bibfield  {title} {\bibinfo {title} {{Monte Carlo eigenvalue and variance estimates from several functional optimizations}},\ }\href@noop {} {\bibfield  {journal} {\bibinfo  {journal} {J. Chem. Phys.}\ }\textbf {\bibinfo {volume} {95}},\ \bibinfo {pages} {6622} (\bibinfo {year} {1991})}\BibitemShut {NoStop}%
\bibitem [{\citenamefont {Kent}\ \emph {et~al.}(1999)\citenamefont {Kent}, \citenamefont {Needs},\ and\ \citenamefont {Rajagopal}}]{Kent1999}%
  \BibitemOpen
  \bibfield  {author} {\bibinfo {author} {\bibfnamefont {P.~R.~C.}\ \bibnamefont {Kent}}, \bibinfo {author} {\bibfnamefont {R.~J.}\ \bibnamefont {Needs}},\ and\ \bibinfo {author} {\bibfnamefont {G.}~\bibnamefont {Rajagopal}},\ }\bibfield  {title} {\bibinfo {title} {{Monte Carlo energy and variance-minimization techniques for optimizing many-body wave functions}},\ }\href@noop {} {\bibfield  {journal} {\bibinfo  {journal} {Phys. Rev. B}\ }\textbf {\bibinfo {volume} {59}},\ \bibinfo {pages} {12344} (\bibinfo {year} {1999})}\BibitemShut {NoStop}%
\bibitem [{\citenamefont {Trail}(2008)}]{Trail2008}%
  \BibitemOpen
  \bibfield  {author} {\bibinfo {author} {\bibfnamefont {J.~R.}\ \bibnamefont {Trail}},\ }\bibfield  {title} {\bibinfo {title} {{Heavy-tailed random error in quantum Monte Carlo}},\ }\href@noop {} {\bibfield  {journal} {\bibinfo  {journal} {Phys. Rev. E}\ }\textbf {\bibinfo {volume} {77}},\ \bibinfo {pages} {016703} (\bibinfo {year} {2008})}\BibitemShut {NoStop}%
\bibitem [{\citenamefont {Snajdr}\ and\ \citenamefont {Rothstein}(2000)}]{Snajdr2000}%
  \BibitemOpen
  \bibfield  {author} {\bibinfo {author} {\bibfnamefont {M.}~\bibnamefont {Snajdr}}\ and\ \bibinfo {author} {\bibfnamefont {S.~M.}\ \bibnamefont {Rothstein}},\ }\bibfield  {title} {\bibinfo {title} {{Are properties derived from variance-optimized wave functions generally more accurate? Monte Carlo study of non-energy-related properties of ${\textrm{H}_2}$, ${\textrm{He}}$, and ${\textrm{LiH}}$}},\ }\href@noop {} {\bibfield  {journal} {\bibinfo  {journal} {J. Chem. Phys.}\ }\textbf {\bibinfo {volume} {112}},\ \bibinfo {pages} {4935} (\bibinfo {year} {2000})}\BibitemShut {NoStop}%
\bibitem [{\citenamefont {Bressanini}\ \emph {et~al.}(2002)\citenamefont {Bressanini}, \citenamefont {Morosi},\ and\ \citenamefont {Mella}}]{Bressanini2002}%
  \BibitemOpen
  \bibfield  {author} {\bibinfo {author} {\bibfnamefont {D.}~\bibnamefont {Bressanini}}, \bibinfo {author} {\bibfnamefont {G.}~\bibnamefont {Morosi}},\ and\ \bibinfo {author} {\bibfnamefont {M.}~\bibnamefont {Mella}},\ }\bibfield  {title} {\bibinfo {title} {{Robust wave function optimization procedures in quantum Monte Carlo methods}},\ }\href@noop {} {\bibfield  {journal} {\bibinfo  {journal} {J. Chem. Phys.}\ }\textbf {\bibinfo {volume} {116}},\ \bibinfo {pages} {5345} (\bibinfo {year} {2002})}\BibitemShut {NoStop}%
\bibitem [{\citenamefont {Sorella}(1998)}]{Sorella1998}%
  \BibitemOpen
  \bibfield  {author} {\bibinfo {author} {\bibfnamefont {S.}~\bibnamefont {Sorella}},\ }\bibfield  {title} {\bibinfo {title} {{Green Function Monte Carlo with Stochastic Reconfiguration}},\ }\href@noop {} {\bibfield  {journal} {\bibinfo  {journal} {Phys. Rev. Lett.}\ }\textbf {\bibinfo {volume} {80}},\ \bibinfo {pages} {4558} (\bibinfo {year} {1998})}\BibitemShut {NoStop}%
\bibitem [{\citenamefont {Lin}\ \emph {et~al.}(2000)\citenamefont {Lin}, \citenamefont {Zhang},\ and\ \citenamefont {Rappe}}]{LinZhangRappe2000}%
  \BibitemOpen
  \bibfield  {author} {\bibinfo {author} {\bibfnamefont {X.}~\bibnamefont {Lin}}, \bibinfo {author} {\bibfnamefont {H.}~\bibnamefont {Zhang}},\ and\ \bibinfo {author} {\bibfnamefont {A.~M.}\ \bibnamefont {Rappe}},\ }\bibfield  {title} {\bibinfo {title} {{Optimization of quantum Monte Carlo wave functions using analytical energy derivatives}},\ }\href@noop {} {\bibfield  {journal} {\bibinfo  {journal} {J. Chem. Phys.}\ }\textbf {\bibinfo {volume} {112}},\ \bibinfo {pages} {2650} (\bibinfo {year} {2000})}\BibitemShut {NoStop}%
\bibitem [{\citenamefont {Umrigar}\ and\ \citenamefont {Filippi}(2005)}]{UmrigarFilippi2005}%
  \BibitemOpen
  \bibfield  {author} {\bibinfo {author} {\bibfnamefont {C.~J.}\ \bibnamefont {Umrigar}}\ and\ \bibinfo {author} {\bibfnamefont {C.}~\bibnamefont {Filippi}},\ }\bibfield  {title} {\bibinfo {title} {{Energy and Variance Optimization of Many-Body Wave Functions}},\ }\href@noop {} {\bibfield  {journal} {\bibinfo  {journal} {Phys. Rev. Lett.}\ }\textbf {\bibinfo {volume} {94}},\ \bibinfo {pages} {150201} (\bibinfo {year} {2005})}\BibitemShut {NoStop}%
\bibitem [{\citenamefont {Cuzzocrea}\ \emph {et~al.}(2020)\citenamefont {Cuzzocrea}, \citenamefont {Scemama}, \citenamefont {Briels}, \citenamefont {Moroni},\ and\ \citenamefont {Filippi}}]{Cuzzocrea2020}%
  \BibitemOpen
  \bibfield  {author} {\bibinfo {author} {\bibfnamefont {A.}~\bibnamefont {Cuzzocrea}}, \bibinfo {author} {\bibfnamefont {A.}~\bibnamefont {Scemama}}, \bibinfo {author} {\bibfnamefont {W.~J.}\ \bibnamefont {Briels}}, \bibinfo {author} {\bibfnamefont {S.}~\bibnamefont {Moroni}},\ and\ \bibinfo {author} {\bibfnamefont {C.}~\bibnamefont {Filippi}},\ }\bibfield  {title} {\bibinfo {title} {{Variational Principles in Quantum Monte Carlo: The Troubled Story of Variance Minimization}},\ }\href@noop {} {\bibfield  {journal} {\bibinfo  {journal} {J. Chem. Theory Comput.}\ }\textbf {\bibinfo {volume} {16}},\ \bibinfo {pages} {4203} (\bibinfo {year} {2020})}\BibitemShut {NoStop}%
\bibitem [{\citenamefont {Grohs}\ and\ \citenamefont {Nobile}(2026)}]{Grohs2026}%
  \BibitemOpen
  \bibfield  {author} {\bibinfo {author} {\bibfnamefont {P.}~\bibnamefont {Grohs}}\ and\ \bibinfo {author} {\bibfnamefont {D.}~\bibnamefont {Nobile}},\ }\bibfield  {title} {\bibinfo {title} {{Is Variational Monte Carlo Robust? Sharp Moment Thresholds and Heavy-Tailed Stochastic Optimization}},\ }\href@noop {} {\bibfield  {journal} {\bibinfo  {journal} {arXiv:2606.26009}\ } (\bibinfo {year} {2026})}\BibitemShut {NoStop}%
\bibitem [{\citenamefont {Sch\"atzle}\ \emph {et~al.}(2023)\citenamefont {Sch\"atzle}, \citenamefont {Szab\'o}, \citenamefont {Mezera}, \citenamefont {Hermann},\ and\ \citenamefont {No\'e}}]{deepqmc2023}%
  \BibitemOpen
  \bibfield  {author} {\bibinfo {author} {\bibfnamefont {Z.}~\bibnamefont {Sch\"atzle}}, \bibinfo {author} {\bibfnamefont {P.~B.}\ \bibnamefont {Szab\'o}}, \bibinfo {author} {\bibfnamefont {M.}~\bibnamefont {Mezera}}, \bibinfo {author} {\bibfnamefont {J.}~\bibnamefont {Hermann}},\ and\ \bibinfo {author} {\bibfnamefont {F.}~\bibnamefont {No\'e}},\ }\bibfield  {title} {\bibinfo {title} {{DeepQMC: An open-source software suite for variational optimization of deep-learning molecular wave functions}},\ }\href@noop {} {\bibfield  {journal} {\bibinfo  {journal} {J. Chem. Phys.}\ }\textbf {\bibinfo {volume} {159}},\ \bibinfo {pages} {094108} (\bibinfo {year} {2023})}\BibitemShut {NoStop}%
\bibitem [{\citenamefont {von Glehn}\ \emph {et~al.}(2023)\citenamefont {von Glehn}, \citenamefont {Spencer},\ and\ \citenamefont {Pfau}}]{vonglehn2023}%
  \BibitemOpen
  \bibfield  {author} {\bibinfo {author} {\bibfnamefont {I.}~\bibnamefont {von Glehn}}, \bibinfo {author} {\bibfnamefont {J.~S.}\ \bibnamefont {Spencer}},\ and\ \bibinfo {author} {\bibfnamefont {D.}~\bibnamefont {Pfau}},\ }\bibfield  {title} {\bibinfo {title} {{A self-attention ansatz for ab-initio quantum chemistry}},\ }in\ \href@noop {} {\emph {\bibinfo {booktitle} {Proc. Int. Conf. Learning Representations (ICLR)}}}\ (\bibinfo {year} {2023})\BibitemShut {NoStop}%
\bibitem [{\citenamefont {Misery}\ \emph {et~al.}(2026)\citenamefont {Misery}, \citenamefont {Gravina}, \citenamefont {Santini},\ and\ \citenamefont {Vicentini}}]{Misery2026}%
  \BibitemOpen
  \bibfield  {author} {\bibinfo {author} {\bibfnamefont {A.}~\bibnamefont {Misery}}, \bibinfo {author} {\bibfnamefont {L.}~\bibnamefont {Gravina}}, \bibinfo {author} {\bibfnamefont {A.}~\bibnamefont {Santini}},\ and\ \bibinfo {author} {\bibfnamefont {F.}~\bibnamefont {Vicentini}},\ }\bibfield  {title} {\bibinfo {title} {{Looking Elsewhere: Improving Variational Monte Carlo Gradients by Importance Sampling}},\ }\href@noop {} {\bibfield  {journal} {\bibinfo  {journal} {Mach. Learn.: Sci. Technol.}\ }\textbf {\bibinfo {volume} {7}},\ \bibinfo {pages} {015035} (\bibinfo {year} {2026})}\BibitemShut {NoStop}%
\bibitem [{\citenamefont {Jin}(2026)}]{Jin2026}%
  \BibitemOpen
  \bibfield  {author} {\bibinfo {author} {\bibfnamefont {D.~Z.}\ \bibnamefont {Jin}},\ }\bibfield  {title} {\bibinfo {title} {{Taming the expressiveness of neural-network wave functions for robust convergence to quantum Many-Body states}},\ }\href@noop {} {\bibfield  {journal} {\bibinfo  {journal} {arXiv:2603.15853}\ } (\bibinfo {year} {2026})}\BibitemShut {NoStop}%
\bibitem [{\citenamefont {Shi}\ \emph {et~al.}(2025)\citenamefont {Shi}, \citenamefont {Cui},\ and\ \citenamefont {Zhou}}]{Shi2025}%
  \BibitemOpen
  \bibfield  {author} {\bibinfo {author} {\bibfnamefont {H.-C.}\ \bibnamefont {Shi}}, \bibinfo {author} {\bibfnamefont {E.-L.}\ \bibnamefont {Cui}},\ and\ \bibinfo {author} {\bibfnamefont {D.}~\bibnamefont {Zhou}},\ }\bibfield  {title} {\bibinfo {title} {{A Variance-Based Convergence Criterion in Neural Variational Monte Carlo for Quantum Systems}},\ }\href@noop {} {\bibfield  {journal} {\bibinfo  {journal} {arXiv:2510.17490}\ } (\bibinfo {year} {2025})}\BibitemShut {NoStop}%
\bibitem [{\citenamefont {Guillemin}\ and\ \citenamefont {Zener}(1929)}]{GuilleminZener1929}%
  \BibitemOpen
  \bibfield  {author} {\bibinfo {author} {\bibfnamefont {V.}~\bibnamefont {Guillemin}, \bibfnamefont {Jr.}}\ and\ \bibinfo {author} {\bibfnamefont {C.}~\bibnamefont {Zener}},\ }\bibfield  {title} {\bibinfo {title} {{Hydrogen-ion wave function}},\ }\href@noop {} {\bibfield  {journal} {\bibinfo  {journal} {Proc. Natl. Acad. Sci. U.S.A.}\ }\textbf {\bibinfo {volume} {15}},\ \bibinfo {pages} {314} (\bibinfo {year} {1929})}\BibitemShut {NoStop}%
\bibitem [{\citenamefont {Kwon}\ \emph {et~al.}(1998)\citenamefont {Kwon}, \citenamefont {Ceperley},\ and\ \citenamefont {Martin}}]{Kwon1998}%
  \BibitemOpen
  \bibfield  {author} {\bibinfo {author} {\bibfnamefont {Y.}~\bibnamefont {Kwon}}, \bibinfo {author} {\bibfnamefont {D.~M.}\ \bibnamefont {Ceperley}},\ and\ \bibinfo {author} {\bibfnamefont {R.~M.}\ \bibnamefont {Martin}},\ }\bibfield  {title} {\bibinfo {title} {{Effects of backflow correlation in the three-dimensional electron gas: Quantum Monte Carlo study}},\ }\href@noop {} {\bibfield  {journal} {\bibinfo  {journal} {Phys. Rev. B}\ }\textbf {\bibinfo {volume} {58}},\ \bibinfo {pages} {6800} (\bibinfo {year} {1998})}\BibitemShut {NoStop}%
\bibitem [{\citenamefont {Kashima}\ and\ \citenamefont {Imada}(2001)}]{Kashima2001}%
  \BibitemOpen
  \bibfield  {author} {\bibinfo {author} {\bibfnamefont {T.}~\bibnamefont {Kashima}}\ and\ \bibinfo {author} {\bibfnamefont {M.}~\bibnamefont {Imada}},\ }\bibfield  {title} {\bibinfo {title} {{Path-Integral Renormalization Group Method for Numerical Study on Ground States of Strongly Correlated Electronic Systems}},\ }\href@noop {} {\bibfield  {journal} {\bibinfo  {journal} {J. Phys. Soc. Jpn.}\ }\textbf {\bibinfo {volume} {70}},\ \bibinfo {pages} {2287} (\bibinfo {year} {2001})}\BibitemShut {NoStop}%
\bibitem [{\citenamefont {Laforgue}(1954)}]{Laforgue1954}%
  \BibitemOpen
  \bibfield  {author} {\bibinfo {author} {\bibfnamefont {A.}~\bibnamefont {Laforgue}},\ }\bibfield  {title} {\bibinfo {title} {{Repr{\'e}sentation de l'erreur par un potentiel et solution it{\'e}rative de l'{\'e}quation de Schr{\"o}dinger}},\ }\href@noop {} {\bibfield  {journal} {\bibinfo  {journal} {C. R. Acad. Sci. Paris}\ }\textbf {\bibinfo {volume} {238}},\ \bibinfo {pages} {1033} (\bibinfo {year} {1954})}\BibitemShut {NoStop}%
\bibitem [{\citenamefont {Preuss}(1964{\natexlab{a}})}]{Preuss1964i}%
  \BibitemOpen
  \bibfield  {author} {\bibinfo {author} {\bibfnamefont {H.}~\bibnamefont {Preuss}},\ }\bibfield  {title} {\bibinfo {title} {{Zur Definition des effektiven Potentials I}},\ }\href@noop {} {\bibfield  {journal} {\bibinfo  {journal} {Theoret. Chim. Acta}\ }\textbf {\bibinfo {volume} {2}},\ \bibinfo {pages} {93} (\bibinfo {year} {1964}{\natexlab{a}})}\BibitemShut {NoStop}%
\bibitem [{\citenamefont {Preuss}(1964{\natexlab{b}})}]{Preuss1964}%
  \BibitemOpen
  \bibfield  {author} {\bibinfo {author} {\bibfnamefont {H.}~\bibnamefont {Preuss}},\ }\bibfield  {title} {\bibinfo {title} {{Zur Definition des effektiven Potentials II}},\ }\href@noop {} {\bibfield  {journal} {\bibinfo  {journal} {Theoret. Chim. Acta}\ }\textbf {\bibinfo {volume} {2}},\ \bibinfo {pages} {98} (\bibinfo {year} {1964}{\natexlab{b}})}\BibitemShut {NoStop}%
\bibitem [{\citenamefont {Turbiner}\ and\ \citenamefont {Guevara}(2007)}]{turbinerguevara2007}%
  \BibitemOpen
  \bibfield  {author} {\bibinfo {author} {\bibfnamefont {A.~V.}\ \bibnamefont {Turbiner}}\ and\ \bibinfo {author} {\bibfnamefont {N.~L.}\ \bibnamefont {Guevara}},\ }\bibfield  {title} {\bibinfo {title} {{A note about the ground state of the hydrogen molecule}},\ }\href@noop {} {\bibfield  {journal} {\bibinfo  {journal} {Collection of Czechoslovak Chemical Communications}\ }\textbf {\bibinfo {volume} {72}},\ \bibinfo {pages} {164} (\bibinfo {year} {2007})}\BibitemShut {NoStop}%
\bibitem [{\citenamefont {Bates}\ \emph {et~al.}(1953)\citenamefont {Bates}, \citenamefont {Ledsham},\ and\ \citenamefont {Stewart}}]{batesledsham1953}%
  \BibitemOpen
  \bibfield  {author} {\bibinfo {author} {\bibfnamefont {D.~R.}\ \bibnamefont {Bates}}, \bibinfo {author} {\bibfnamefont {K.}~\bibnamefont {Ledsham}},\ and\ \bibinfo {author} {\bibfnamefont {A.~L.}\ \bibnamefont {Stewart}},\ }\bibfield  {title} {\bibinfo {title} {{Wave functions of the hydrogen molecular ion}},\ }\href@noop {} {\bibfield  {journal} {\bibinfo  {journal} {Philos. Trans. R. Soc. London A}\ }\textbf {\bibinfo {volume} {246}},\ \bibinfo {pages} {215} (\bibinfo {year} {1953})}\BibitemShut {NoStop}%
\end{thebibliography}%

\appendix
\section{The Choice of Centering Point \texorpdfstring{$E_R$}{ER} and the Gradient Calculation}
\label{app:centering}

The mean uniquely minimizes the expectation value $\mathbb{E}[(X-E_R)^2]$ over $E_R$, while the median uniquely minimizes $\mathbb{E}[|X-E_R|]$. Variance minimization is thus self-centering by construction, but an $L^1$-type dispersion measure has no such built-in anchor unless explicitly centered at the median. Three choices of $E_R$ are useful in practice:

\begin{description}
\item[\textbf{MEAN}] $E_R=\langle E_L\rangle$, the common choice, recomputed at every optimization step;
\item[\textbf{MEDIAN}] $E_R=\mathrm{med}(E_L)$, the $L^1$ analogue of the mean, in the precise sense above;
\item[\textbf{EFIXED}] $E_R=$ an externally supplied constant fixed for the entire run.
\end{description}

Conroy's 1964 argument against MEAN, already noted above, is EFIXED's historical motivation.
Cuzzocrea and coworkers showed, however, that even a reference initialized at the energy of the target state does not prevent the optimization from drifting to states of lower variance \cite{Cuzzocrea2020}: a fixed reference stabilizes the gradient, but cannot create minima that the variance landscape lacks.

Centering an $L^1$-type loss on the mean is only partially robust: the absolute value mitigates the impact of outliers, but the mean itself is still dragged around by the very configurations one hoped to be insensitive to. The median is robust in both respects: for a heavy-tailed local energy, the sample mean's variance grows large or is formally undefined (in the extreme case, a Cauchy-distributed local energy yields an infinite variance for the sample mean), while the sample median's asymptotic variance stays finite.

Consider $g(\delta)$, the loss function assigned to configuration $\mathbf{R}_m$, to be $\delta^2,|\delta|,\log(1+\delta^2/s^2)$, or
 $\log(1+\delta^4/s^4)$, where $\delta_m = (E_L(\mathbf{R}_m)-E_R)$. 
With $w_m$ the weight of configuration $m$, the cost function is $\mathcal{L}_g = W^{-1}\sum_m w_m\,g(\delta_m)$ and $W=\sum_m w_m$.
MEAN, MEDIAN, and EFIXED are three choices for the $E_R$ entering both $\delta_m$ and, implicitly, $\mathcal{L}_g$. Its derivative with respect to a variational parameter $q_j$ is:

\begin{align}
\frac{d\mathcal{L}_g}{dq_j} = & \underbrace{\frac{1}{W}\sum_m w_m\,g'(\delta_m)\,\frac{dE_{L,m}}{dq_j}}_{\text{(A)}} \nonumber \\
& + \underbrace{\frac{2}{W}\sum_m w_m\,\frac{d\log\psi_m}{dq_j}\Big[g(\delta_m)-\mathcal{L}_g\Big]}_{\text{(B): reweighting}} \nonumber \\
& - \underbrace{\frac{1}{W}\Big(\sum_m w_m\,g'(\delta_m)\Big)\frac{dE_R}{dq_j}}_{\text{(C): centering correction}} \, .
\end{align}

Terms (A) and (B) are the ordinary local energy and reweighting derivatives common to every VMC scheme; ignoring term (C) introduces a systematic gradient bias. Term (C) vanishes only when
 $\sum_m w_m\,g'(\delta_m)=0$ at the chosen $E_R$. For variance ($g=\delta^2$) with $E_R=\langle E \rangle$, this holds as an exact
identity for any sample, by definition.

It is tempting to assume that the same holds for MAD ($g=|\delta|$) if we choose $E_R$ to be the median, by the same argument. This is exact for a continuous distribution, but not for the finite, weighted sample of an actual Monte Carlo run: the configuration achieving the (weighted) median carries a finite, not infinitesimal, share of the total weight, so $\sum_m w_m\,\mathrm{sign}(E_{L,m}-E_R)$ at that configuration's own
local energy differs from zero by an amount proportional to its weight.

Despite this, the correction introduces negligible computational overhead: the (weighted) median of a discrete sample is always the local energy of one particular configuration $m^*$, implying $dE_R/dq_j=dE_{L,m^*}/dq_j$, which is an ordinary derivative already needed elsewhere in the calculation. Term (C) is thus computed exactly for MEDIAN as well as MEAN, and the general formula is exact for every loss function/centering combination used here.
EFIXED needs no correction at all, since $E_R$ is constant and $dE_R/dq_j\equiv
0$. For the $\mathcal{L}_{\mathrm{C}}$ and $\mathcal{L}_{-4}$ functionals, the exact centering is a self-consistent weighted mean whose weights redescend to zero for outliers; to avoid an iterative numerical solution at every step, we approximate it with the sample median instead, preserving the outlier suppression while keeping the gradient exact via that one walker's derivative.

\end{document}